\documentclass[aps,prc,twocolumn,superscriptaddress,showpacs,floatfix]{revtex4}
\usepackage{graphicx,amsmath,amssymb,bm}

\newcommand{\beq}{\begin{equation}}
\newcommand{\eeq}{\end{equation}}
\newcommand{\bea}{\begin{eqnarray}}
\newcommand{\eea}{\end{eqnarray}}
\newcommand{\ben}{\begin{eqnarray*}}
\newcommand{\een}{\end{eqnarray*}}

\newcommand{\simge}{\hspace*{0.2em}\raisebox{0.5ex}{$>$}
     \hspace{-0.8em}\raisebox{-0.3em}{$\sim$}\hspace*{0.2em}}

\newcommand{\boldtau}{\mbox{\boldmath $\tau$}}

\begin{document}

\title{Renormalization of One-Pion Exchange 
and Power Counting}
\author{A. Nogga}
\email[E-mail:~]{a.nogga@fz-juelich.de}
\affiliation{Institut f\"ur Kernphysik, Forschungszentrum J\"ulich, Germany}
\author{R.G.E. Timmermans}
\email[E-mail:~]{timmermans@kvi.nl}
\affiliation{Theory Group, KVI, University of Groningen, The Netherlands}
\author{U. van Kolck}
\email[E-mail:~]{vankolck@physics.arizona.edu}
\affiliation{Department of Physics, University of Arizona, Tucson, 
AZ 85721, USA }

\begin{abstract}
The renormalization of the chiral nuclear interactions is studied. 
In leading order, the cutoff dependence is related 
to the singular tensor interaction of the one-pion exchange potential. 
In S waves and in higher partial waves where the tensor force is repulsive 
this cutoff dependence can be absorbed by counterterms expected at that order. 
In the other partial waves additional contact interactions are necessary.
The implications of this finding for the effective-field-theory
program in nuclear physics are discussed.
\end{abstract}

\maketitle

\section{Introduction}
\label{sec:intro}

It is commonly accepted that QCD is the correct theory for the 
strong interaction in the energy regime of interest for 
nuclear physics. At the same time, this is of limited practical value, because 
in this energy regime QCD needs to be solved  non-perturbatively. Lattice 
simulations, which are in principle able to deal with this non-perturbative
character are, for systems with $A \ge 2$ nucleons, still in their beginning
stages \cite{fukugita95,beane03a}.

A possible way out of this dilemma is the application of 
chiral perturbation theory (ChPT) 
to nuclear systems \cite{kolck99,bedaque02a}. 
Making use of the spontaneously-broken 
chiral symmetry of QCD, one can formulate an effective field theory (EFT) 
involving nucleons ($N$) and the Goldstone bosons related to 
chiral-symmetry breaking, the pions ($\pi$). ChPT is a powerful 
approach because it relates processes with different numbers of pions. 
In the purely-pionic and one-nucleon sectors, 
the Goldstone-boson character of pions guarantees that
amplitudes can be expanded in powers of momenta \cite{weinberg79,bernard95}.
In the few-nucleon sector, however, the 
existence of bound states clearly shows the non-perturbative 
nature of the problem. 

Weinberg \cite{weinberg90,weinberg91} recognized that this is caused  by 
an infrared enhancement in the propagation of two or more nucleons. 
He suggested that the calculation of a generic nuclear amplitude
should consist of two steps.
In the first step, 
one defines the
nuclear potential as the sum of 
``irreducible'' sub-diagrams that do not contain purely-nucleonic
intermediate states,
and truncates the sum according to
a simple extension of the standard ChPT power counting.
In a second step, the potential is 
iterated to all orders, which can be done by using the Lippmann-Schwinger (LS) 
or Schr\"odinger equations.
The potential includes pion exchanges and contact interactions,
which represent the contributions of more massive degrees of freedom.
Assuming that contact interactions 
obey naive dimensional analysis,
only a finite number of pion exchanges and contact interactions
contribute to the potential at any given order: 
two non-derivative two-nucleon contact interactions
and one-pion exchange (OPE) in leading order, 
and derivative contact interactions,
two-pion exchange (TPE) and more-pion exchanges, and few-nucleon interactions
in subleading orders.

This power counting naturally explains 
that two-nucleon interactions are more important than three-nucleon 
interactions, {\it etc.} \cite{weinberg92}.
The resulting two-nucleon ($N\!N$) \cite{ordonez92,kaiser02} 
and three-nucleon ($3N$)  \cite{kolck94,friar99a} potentials
provide a quantitative description of
few-nucleon systems
\cite{ordonez94,ordonez96,epelbaum00,entem02b,entem02a,entem03a,epelbaum05a,epelbaum01a,epelbaum02a,epelbaum02c}.
In addition, this approach matches well with the Nijmegen energy-dependent
partial-wave analysis (PWA) of $N\!N$ scattering data \cite{stoks93a}. 
In this analysis, a separation of long- and short-distance physics is
implemented by solving the partial-wave
Schr\"odinger equation with a long-range potential
that consists of OPE and TPE (and the electromagnetic interaction),
and a boundary condition with as many short-range parameters as are needed for
an optimal description of the observables. The pion mass and OPE parameters
\cite{stoks93b,kolck98}
and even TPE parameters \cite{rentmeester99,rentmeester03}
could be determined from the $N\!N$ scattering data, in good agreement with
values obtained from pion-nucleon scattering \cite{bernard97,fettes99,buttiker00}.
 
However, Weinberg's power counting has been criticized. 
As in any EFT,
a regularization procedure is required in order to separate
high- and low-energy physics.
Since this separation is arbitrary,
a consistent power counting 
should provide sufficient counterterms at each order 
to absorb any cutoff dependence in the limit of large cutoffs.
Because the solution of the LS equation is numerical in character,
an explicit check of cutoff independence
is challenging. 
This led Kaplan and coworkers \cite{kaplan96,kaplan98a,kaplan98b} 
to examine a few of the
diagrams 
contributing to the $N\!N$ $T$ matrix.
They identified in two-loop diagrams ultraviolet divergences 
proportional to the square of the pion mass 
and of the external momenta,
which are
present in leading order
but
cannot be absorbed by the available counterterms.
They concluded 
that pion exchange should not be fully iterated,
but instead be treated in finite order in perturbation theory. 
Quantitative calculations at higher order showed, however, 
that this idea fails in some partial waves 
at momenta comparable to the pion mass \cite{fleming00}. 

For smaller momenta, one can
integrate out the pion and construct a ``pionless'' EFT,
which is very successful within its limited range \cite{bedaque02a}.
Nevertheless, a lot of interesting nuclear physics is thought
to take place at momenta of the order of the pion mass.
(The Fermi momentum of isospin-symmetric nuclear matter,
for example, is about 300 MeV.)
It seems unavoidable
that in this larger momentum range pion
exchange has to be iterated.

It is now well known that the renormalization
of an EFT is not necessarily the same as that of its
perturbative series. 
This is seen clearly in the three-body problem in
the pionless EFT \cite{bedaque99b,bedaque99a,bedaque00}. 
The origin of this feature lies in the renormalization
of singular potentials \cite{frank71,beane01}.
In the specific case of OPE, the singularity is
the $1/r^3$ 
behavior of the tensor force 
in spin-triplet channels. 
It has been found that 
the cutoff dependence of an uncoupled  $1/r^n$ interaction
in the S wave
can be absorbed into one counterterm \cite{beane01}. 

The renormalization of OPE in lowest waves was reexamined from
the non-perturbative viewpoint in Refs. \cite{frederico99,beane02}. 
The problem with the ultraviolet divergence proportional
to the pion mass squared in the $^1$S$_0$ channel 
persists in this context \cite{beane02}.
On the other hand,
the divergence associated with momenta, present in the
$^3$S$_1$-$^3$D$_1$ coupled channel,
{\it can} be absorbed into the available leading-order counterterm
\cite{frederico99,beane02}. 
With a further expansion around the chiral limit, 
Weinberg's power counting seems to be consistent in 
a {\it non-perturbative} calculation 
of these waves \cite{beane02,valderrama04,valderrama05}.
(For a different conclusion, see Ref. \cite{eiras03}.)
 
OPE contributes, however, also in higher partial waves.
The naive power counting does not 
predict leading-order counterterms in these partial waves. 
However, the singularity of 
the tensor interaction exists in all the spin-triplet channels.
In fact, it has been argued that for an uncoupled singular interaction
boundary conditions need to be fixed in all 
waves where the potential is attractive \cite{behncke68}.
Therefore, cutoff dependence can be expected in some spin-triplet channels
if there are no corresponding counterterms,
posing a significant difficulty for Weinberg's power counting.

Another important renormalization issue concerns few-nucleon forces.
In the pionless theory it has been shown that 
consistent renormalization requires a $3N$ force in leading order 
\cite{bedaque00}. This result is not necessarily in contradiction
with Weinberg's power counting in the ``pion\-ful'' EFT,
because the $3N$ force in the pionless theory includes contributions
that  are iterations of the $N\!N$ force with
intermediate-state nucleons of momentum ${\cal O}(m_\pi)$ in the 
pion\-ful theory.
The two EFTs have $N\!N$ interactions with different ultraviolet
behaviors. Whether OPE sufficiently softens the
asymptotic behavior of the $3N$ LS equation is an
issue that remains unresolved. 

In practice, the renormalization issue has been sidestepped 
by choosing  
rather low cutoffs to regularize 
the LS equation and by varying the cutoffs only in a very limited range 
\cite{ordonez94,ordonez96, epelbaum00,entem02b,entem02a,entem03a,epelbaum05a,epelbaum01a,epelbaum02a,epelbaum02c}. 
Cutoff dependence has generally been observed 
in higher partial waves; see, for example, 
the discussion in Ref. \cite{meissner01a} regarding higher orders
in the Weinberg expansion.
For relatively small cutoff variations,
it has been noticed that the resulting variations in the phase
shifts decreases with increasing order.
It has then been assumed that the observed cutoff dependence 
is of the order of the error expected from the truncation of 
the expansion. 
In fact, it has been argued that the EFT involving nucleons 
and pions necessarily involves a mild cutoff dependence and that 
cutoff values exist that are optimal for the convergence of the expansion
\cite{lepage97,gegelia04}. 

An in-depth study of the cutoff dependence in higher $N\!N$ partial waves
and in the $3N$ system
still needs to be performed.
This study is the aim of this work.
We seek
to quantify the cutoff dependence in lowest order and, if possible,
to identify ranges of cutoffs 
in which only small variations of observables 
occur. We then discuss how
to absorb the cutoff dependence in a finite number of counterterms. 
We first consider $N\!N$ scattering, in which case we compare our
results with the phase shifts and mixing angles from the energy-dependent
Nijmegen PWA93 \cite{stoks93a}, which provide an optimal representation
of the $N\!N$ database.
(The fact that this PWA93 did not yet include TPE \cite{rentmeester99,
rentmeester03} does not affect our investigation of OPE.) 
We then extend our analysis to the $3N$ bound state. 
We will restrict ourselves to total 
$N\!N$ angular momentum $j \le 4$, which 
is sufficient to study the $3N$ binding energy. 
As we are going to show, our results have significant implications:
we present here a modification of power counting that is consistent
with all known results, and could become a new basis to
organize nuclear interactions in the EFT with pions.

Section~\ref{sec:numerics} describes the interaction and our approach 
to regularize and solve the $N\!N$ 
Lippmann-Schwinger equation. 
In Section~\ref{sec:res},
we identify problematic partial waves, explicitly show their cutoff 
dependence, and present
counterterms that generate cutoff-independent phase shifts
in reasonable agreement with the PWA.
Section~\ref{sec:tri} is devoted to the  
$3N$ system. 
The implications of our findings to power counting in nuclear ChPT
are analyzed in Section~\ref{sec:counting}. Finally,
our conclusions and an outlook are given in Section~\ref{sec:concl}.

\section{Regularization of the Lippmann-Schwinger equation}
\label{sec:numerics}

We first consider $N\!N$ scattering in the center-of-mass frame.
We denote by $\mu$ the reduced mass ($m_N= 2\mu$ is the nucleon mass),
by $E$ the energy,
and by $\vec p$ and $\vec p\,'$ the relative momenta 
before and after interaction; the momentum transfer is
$\vec q = \vec p - \vec p\,'$. 
The relative distance between the two nucleons is
$\vec r$.
The standard Pauli matrices in spin and isospin space
are denoted by  $\vec \sigma_i$ and 
${\boldtau}_i$,
respectively. 

With a standard normalization for plane waves,
the LS equation for the $T$ matrix reads, in momentum space,
\begin{eqnarray}
& & T(\vec p\,',\vec p,E)  =  V(\vec p\,',\vec p) \cr 
& & \quad +  \int d^3p''  V(\vec p\,',\vec p\,'')
 \frac{1}{E + i \epsilon - \frac{{\vec p\,''}^2}{2 \mu} }  
 \ T(\vec p\,'',\vec p,E),
\end{eqnarray}
with $V$ the potential. 
The OPE potential is
\begin{equation}
V_{1\pi}(\vec q\,) = - \frac{1}{(2\pi)^3 } 
\left( \frac{g_A }{2 f_\pi} \right)^2 {\boldtau}_1 \cdot {\boldtau}_2 
\frac{ ( \vec \sigma_1 \cdot \vec q\,) 
\ ( \vec \sigma_2 \cdot \vec q\,)}{ {\vec q}^{\,2} + m_\pi^2 },   
\end{equation} 
where $m_\pi$ is the pion mass. 
In lowest order the strength of OPE
is completely  determined
by the axial-coupling constant $g_A = 1.26$ and the 
pion-decay constant $f_\pi =92.4$~MeV.  

In addition to pion exchanges, the EFT contains
short-range interactions that represent high-energy
degrees of freedom that have been integrated out.
The simplest are two contact interactions
\begin{equation}
V_c = \frac{1}{ 4 \pi} \ \frac{1}{(2\pi)^3 } 
      \left( c_s \ P_s + c_t \ P_t  \right),
\end{equation}
where 
we used the projectors onto spin-triplet 
and spin-singlet states, $P_t$ and $P_s$. 
The two strength parameters
$c_s$ and $c_t$ need to be 
determined from $N\!N$ scattering data,
for instance from the scattering lengths
in the $^1$S$_0$ and $^3$S$_1$ channels.
It is possible to write
\begin{equation}
c_s = C_0 + m_\pi^2 D_2 +\ldots, \label{C0D2}
\end{equation}
where the parameters $C_0$ and $D_2$ are independent of
the quark masses.

For the numerical solution of the LS equation, we need to 
introduce a regulator $f(p',p )$ 
that effectively cuts momenta at a cutoff $\Lambda$. 
The regularization procedure is an arbitrary splitting of short-range physics
into the high-momentum region of loops and contact interactions.
Low-energy physics should, of course,
be independent of the choice of regulator (renormalization-group
invariance),
once the dependence of contact parameters on the cutoff 
is taken into account. 
It is convenient  for the partial-wave decomposition 
to perform the regularization using momentum cutoff functions 
depending on  $\vec p$ and $\vec p\,'$ rather than on $\vec q$.
Here we use
\begin{equation}
f(p',p ) = e^{-  (p^4 + p'^4)/\Lambda ^4}.
\end{equation}
This leads to nonlocal interactions in configuration space.
However,  because 
the regulator only depends on the magnitude of the relative momenta, 
it does not influence the partial-wave decomposition. 
This 
guarantees that contact interactions act in specific partial waves, 
independent of $\Lambda$. 
In particular, it
implies that $V_{c}$ only acts in the two S waves. 

For the following discussion, it is useful to look also at the 
configuration space expression for OPE,
\begin{equation}
\label{eq:1piconf1}
V_{1\pi}(\vec r\,) = \frac{ m_\pi ^3 }{ 12 \pi } 
\left(\frac{ g_A  }{ 2 f_\pi} \right)^2
{\boldtau}_1 \cdot {\boldtau}_2 
\left[ T(r) \ S_{12} + Y(r) \ \vec \sigma_1 \cdot \vec \sigma_2 \right], 
\end{equation}
where
\begin{eqnarray}
\label{eq:1piconf2}
 T(r) & = & \frac{ e^{- m_\pi r } }{ m_\pi r} \ 
            \left[ 1 + \frac{3}{ m_\pi r } 
                   + \frac{ 3}{ (m_\pi r)^2 } \right], \cr
 Y(r) & = &   \frac{ e^{- m_\pi r }}{ m_\pi r},
\end{eqnarray}
and the tensor operator is
\begin{equation}
\label{eq:1piconf3}
 S_{12} = 3 (\vec \sigma_1 \cdot \hat r) ( \vec \sigma_2 \cdot \hat r) 
                         -  \vec \sigma_1 \cdot \vec \sigma_2. 
\end{equation}
The tensor force $T(r)$ of OPE contains a singular interaction $\sim 1/r^3$
that acts in the spin-triplet waves; the tensor force is zero in the
spin-singlet channels. Using the partial-wave matrix elements given
in Table~\ref{tab:tensor} one can identify whether the tensor force
is attractive or repulsive in specific partial waves. This we will
require below.

\begin{table}
\begin{tabular}{l||l|ccc}
$t$    & $s=1$ & $l=j-1$ &   $l=j$ &   $l=j+1$ \cr
\hline\hline
$t=1$ & $l'=j-1$&   $-2 \ \frac{ j-1}{ 2j+1}$  &  0   &  $6 \frac{ \sqrt{j(j+1)}}{ 2j+1 }$ \cr
      & $l'=j $ &   0                                   &  2   &               0                            \cr
      & $l'=j+1$& $6 \frac{ \sqrt{j(j+1)}}{ 2j+1 } $& 0 &  $-2 \ \frac{ j+2}{2j+1}$ \cr
\hline
$t=0$ & $l'=j-1$&   $6 \ \frac{ j-1}{ 2j+1}$  &  0   &  $-18 \frac{ \sqrt{j(j+1)}}{ 2j+1 } $\cr
      & $l'=j$  &   0                                   &  $-6$   &               0                            \cr
      & $l'=j+1$& $-18 \frac{ \sqrt{j(j+1)}}{ 2j+1 } $& 0 &  $6 \ \frac{ j+2}{2j+1}$ \cr
\end{tabular}
\caption{\label{tab:tensor} 
Matrix elements of the operator ${\boldtau}_1 \cdot {\boldtau}_2\,S_{12}$
           for spin-triplet channels with total angular momentum $j$.
           The matrix elements depend on the isospin $t$,
              and on the incoming and outgoing 
              angular momentum $l$ and $l'$.}
\end{table}

\section{Nucleon-nucleon phase shifts}
\label{sec:res}

Our aim is to study the dependence of observables on the chosen value 
for the cutoff $\Lambda$. 
We have performed a partial-wave decomposition of the interaction 
described in the previous section and then 
solved the LS equation and extracted phase shifts. 
The explicit expressions are 
summarized in Appendix~\ref{app:onepi}.
We study the cutoff dependence of the phase shifts
in leading order (LO), or ${\cal O}(Q^0)$, in Weinberg's power counting.
We consider $\Lambda$ in a wide range, between 2~fm$^{-1}$ and 
20~fm$^{-1}$. 

\begin{figure*}[btp]
\begin{center}
\includegraphics[width=15cm]{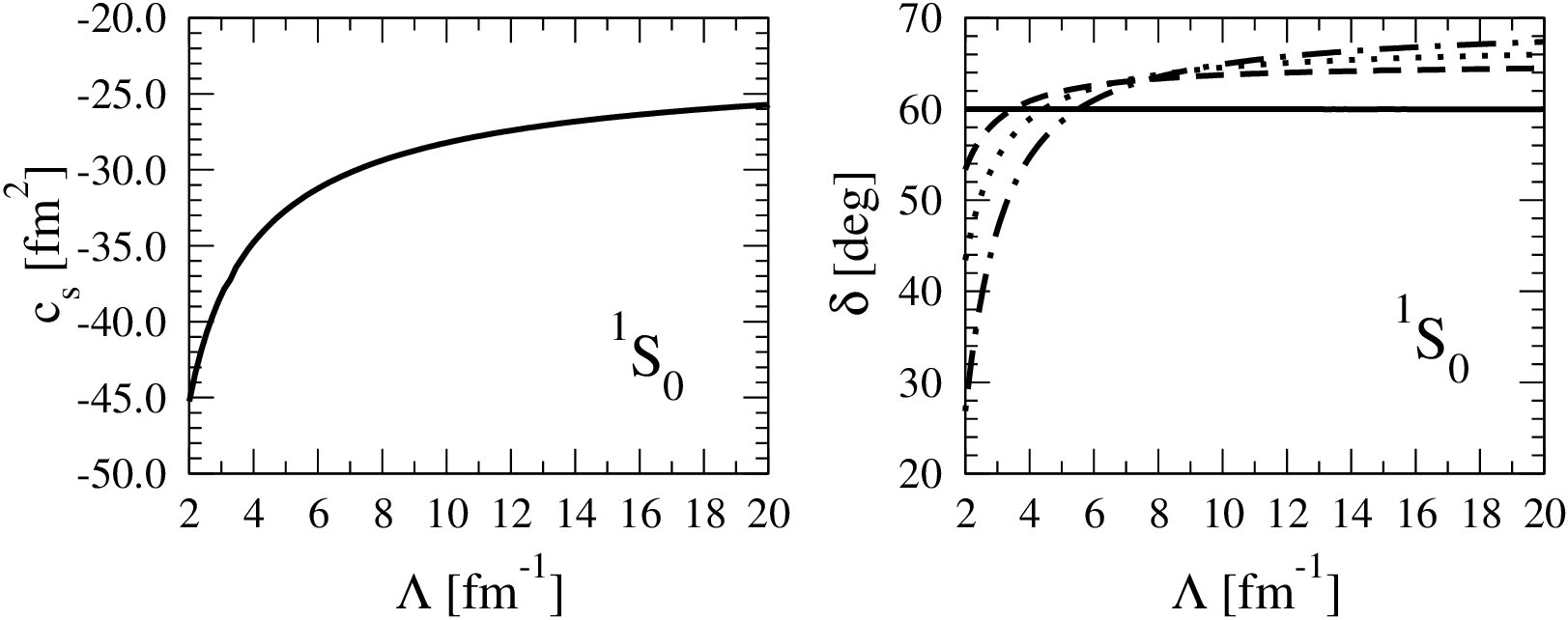}
\end{center}
\vspace{-0.6cm}
\caption{\label{fig:lamdep1s0} Fit result for the counterterm $c_s$ 
as a function of the cutoff,
and the resulting  
cutoff dependence of the $^1$S$_0$ phase
shifts at laboratory energies of 10~MeV (solid line), 50~MeV (dashed line), 
100~MeV (dotted line),
and 190~MeV (dash-dotted line). }
\end{figure*}

\begin{figure*}[btp]
\begin{center}
\includegraphics[width=15cm]{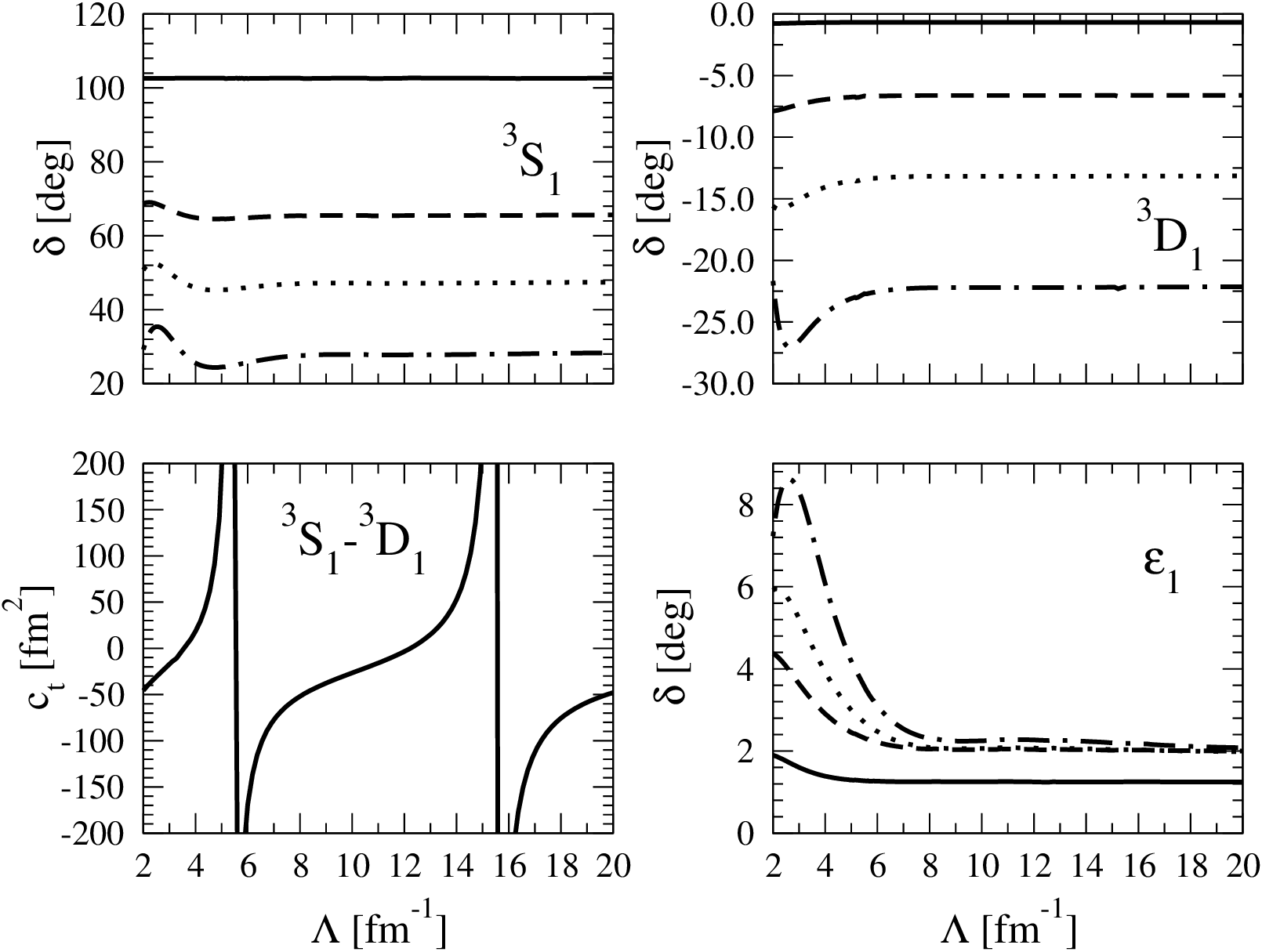}
\end{center}
\vspace{-0.6cm}
\caption{\label{fig:lamdep3s1} Fit result for the counterterm $c_t$ 
as a function of the cutoff,
and the resulting  
cutoff dependence of the $^3$S$_1$-$^3$D$_1$ phase shifts and the
mixing angle $\varepsilon_1$
at laboratory energies of 10~MeV (solid line), 50~MeV (dashed line), 
100~MeV (dotted line),
and 190~MeV (dash-dotted line). }
\end{figure*}

We start with the S-wave channels, which were previously
examined in Refs. \cite{frederico99,beane02,valderrama04,valderrama05} 
with different regularizations. 
We fit $c_s$ and $c_t$ 
to the  $^1$S$_0$  and $^3$S$_1$
phase shifts at 10~MeV and we confirm the cutoff independence found 
in Refs. \cite{frederico99,beane02}, 
as can been seen in Figs.~\ref{fig:lamdep1s0}  
and \ref{fig:lamdep3s1}. 
In Fig.~\ref{fig:lamdep1s0} we show the running of  $c_s$ 
with the cutoff $\Lambda$,
and the resulting cutoff 
dependence of the $^1$S$_0$ phase shifts at various laboratory energies.
In Fig.~\ref{fig:lamdep3s1} we show the 
corresponding results for  $c_t$, and the 
$^3$S$_1$ and $^3$D$_1$ phase shifts and the mixing angle $\varepsilon_1$.
One sees that the cutoff dependence of the phase shifts is small
for $\Lambda\simge 5$ fm$^{-1}$, but it increases for increasing energy,
as expected in an EFT.
It is interesting to note that
 $c_t(\Lambda)$ displays a nice limit-cycle-like behavior,
similar to the $3N$ force in the $3N$ 
problem in pionless EFT \cite{bedaque00}, which is solved using 
a regulator similar to ours. 
Since the counterterm strength behaves 
differently in Ref. \cite{beane02}, 
where a coordinate-space regulator was employed instead,
we conclude that this behavior is 
regulator dependent. 
This is in line with a similar recent 
finding for a central potential \cite{braaten04a,hammer05}.
Note that the running found here is similar to that observed in 
Ref.~\cite{epelbaum03b} for 
a counterterm in a different channel at higher orders but with  
the same regularization of the LS equation. This suggests that the form 
of the running is perhaps more influenced by the regulator than the specific 
form of the singularity of the interaction.

\begin{figure*}[btp]
\begin{center}
\includegraphics[width=15cm]{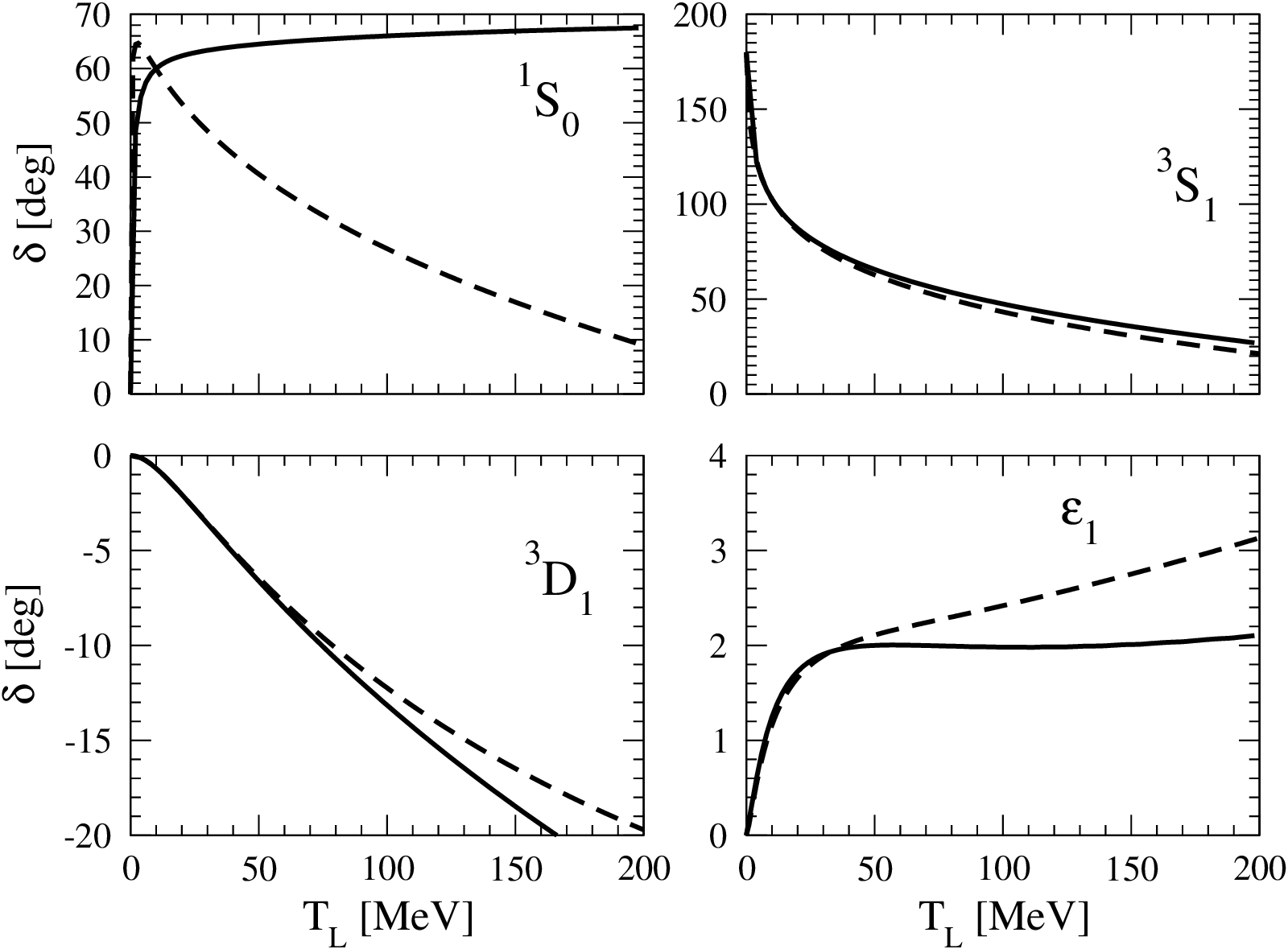}
\end{center}
\vspace{-0.6cm}
\caption{\label{fig:phaseswave} Comparison of 
the $^1$S$_0$ and $^3$S$_1$-$^3$D$_1$ phase shifts 
and the mixing angle $\varepsilon_1$
(as function of the laboratory energy)
in lowest order 
for  $\Lambda=20$~fm$^{-1}$ (solid line)
to the Nijmegen PWA (dashed line).}
\end{figure*}

The resulting phase shifts as function of
the laboratory energy are
shown in Fig.~\ref{fig:phaseswave}.
In the $^1$S$_0$ channel, we recover the known strong deviation
of the LO result from the PWA.
This is related to the relatively-large effective-range parameter 
in this partial wave \cite{kaplan99b,beane02}, which 
cannot be reproduced without a two-derivative contact interaction.
This problem is solved
once the latter is included in subleading order
(see, {\it e.g.}, Ref. \cite{beane02}). In the coupled 
$^3$S$_1$-$^3$D$_1$ channels, we find an encouraging agreement 
with the PWA.
The mixing angle $\varepsilon_1$ is underpredicted, when
one goes to the limit of high $\Lambda$. In this limit the agreement 
with the PWA has, however, 
improved compared to the 
choice $\Lambda\simeq 3$ fm$^{-1}$
used in the literature  \cite{epelbaum00}.

Despite these positive results,
there are potential problems in other waves. 
The explicit expression in Eq.~(\ref{eq:1piconf1})  
for $V_{1\pi}$ in configuration space suggests
that, 
because of the singularity of $T(r)$,
we can expect a number of bound states 
---infinite in the limit $\Lambda \rightarrow \infty$---
in all channels where the tensor force is attractive.
A consequence would be 
cutoff dependence in these waves.
We will, therefore, study higher partial waves, starting from 
P waves, since these, according to Weinberg's power counting, 
should not require counterterms
in lowest order. 

\begin{figure*}[btp]
\begin{center}
\includegraphics[width=15cm]{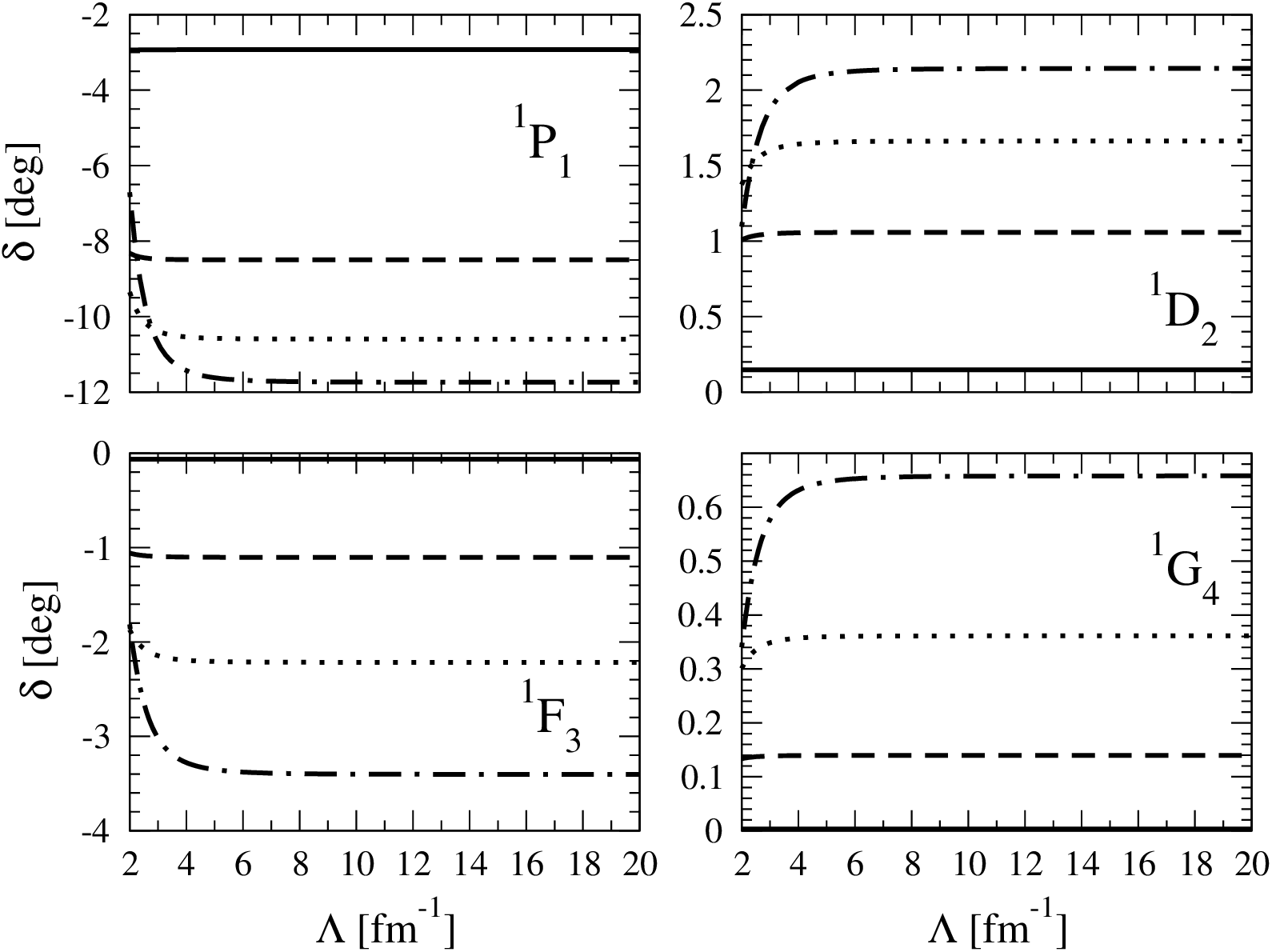}
\end{center}
\vspace{-0.6cm}
\caption{\label{fig:lamdepsing}  
Cutoff dependence of the singlet phase shifts 
for various partial waves. Results are given for laboratory energies of 
10~MeV (solid line), 50~MeV (dashed line), 100~MeV (dotted line), and 190~MeV 
(dash-dotted line).}
\end{figure*}

\begin{figure*}[btp]
\begin{center}
\includegraphics[width=15cm]{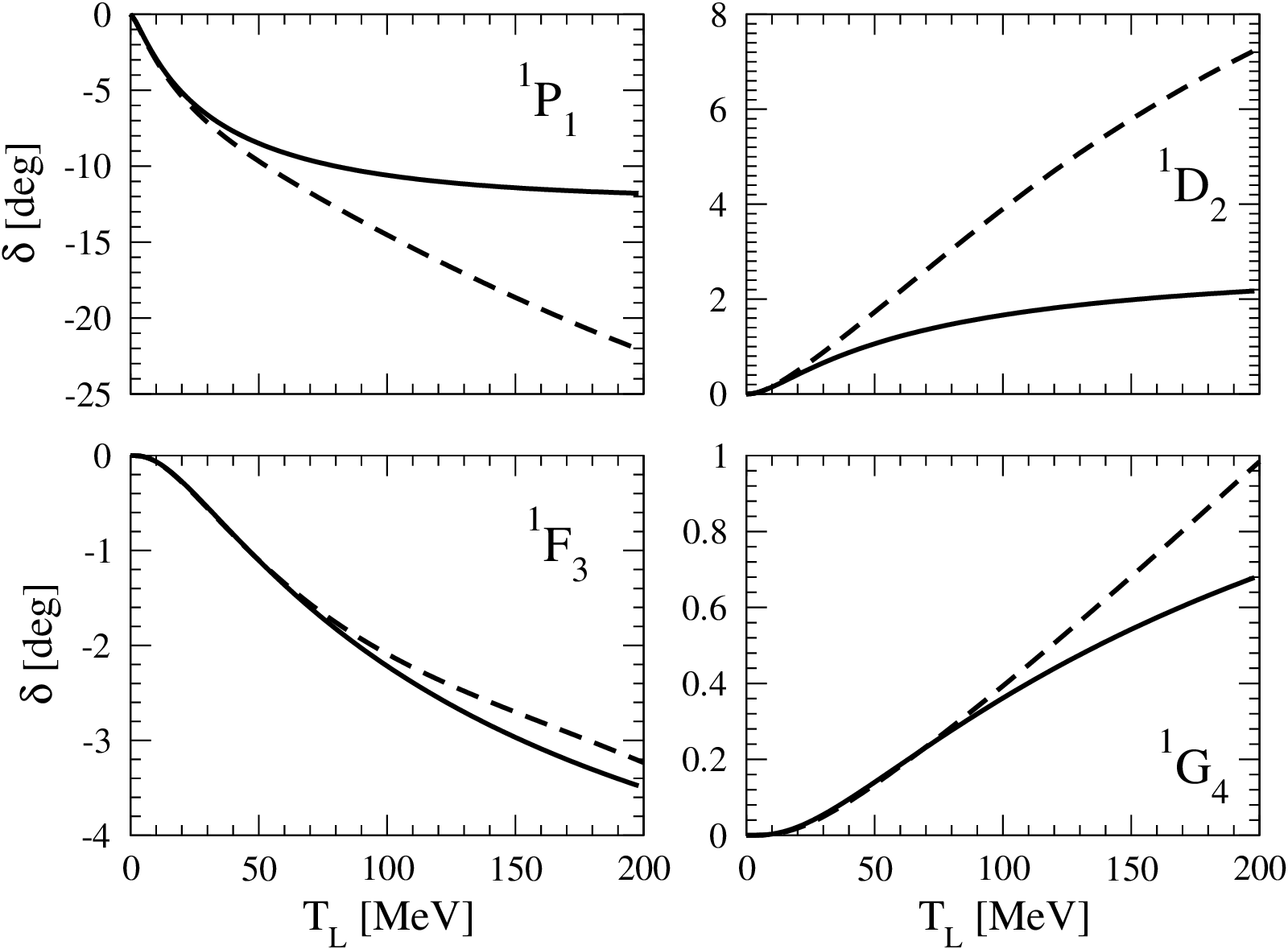}
\end{center}
\vspace{-0.6cm}
\caption{\label{fig:exptsing}  Comparison of the LO singlet 
phase shifts (as function of the laboratory energy)
for $\Lambda=20$~fm$^{-1}$ (solid line)
to the Nijmegen PWA (dashed line). }
\end{figure*}

We start with 
spin-singlet channels, 
where the tensor interaction is zero and the potential is non-singular. 
The results for the $^1$P$_1$,  $^1$D$_2$,  $^1$F$_3$, and
$^1$G$_4$ phase shifts 
as function of $\Lambda$ at four different lab
energies are shown in Fig.~\ref{fig:lamdepsing}. 
It is seen that indeed the dependence on $\Lambda$ becomes smaller
with increasing $\Lambda$. 
Even for a rather high energy of 190~MeV, we find in all 
cases only negligible changes in the phase shifts for 
$\Lambda \simge 5$~fm$^{-1}$. 
This supports the claim that in these channels no inconsistency in 
Weinberg's power counting 
exists. Fig.~\ref{fig:exptsing} compares the resulting phase shifts
to the PWA. For the P and D waves 
the agreement is good below 30~MeV. Above that energy 
significant higher-order contributions are necessary 
to improve agreement with the PWA. For the F and G waves,
where contact interactions are expected at even higher orders, 
the agreement is much better for energies up to 100~MeV. 
We note that for the energies below 30-50~MeV, where all singlet phase shifts  
are described well by the lowest-order predictions, the cutoff dependence 
is already negligible for $\Lambda \simge 2$~fm$^{-1}$. 

\begin{figure*}[btp]
\begin{center}
\includegraphics[width=15cm]{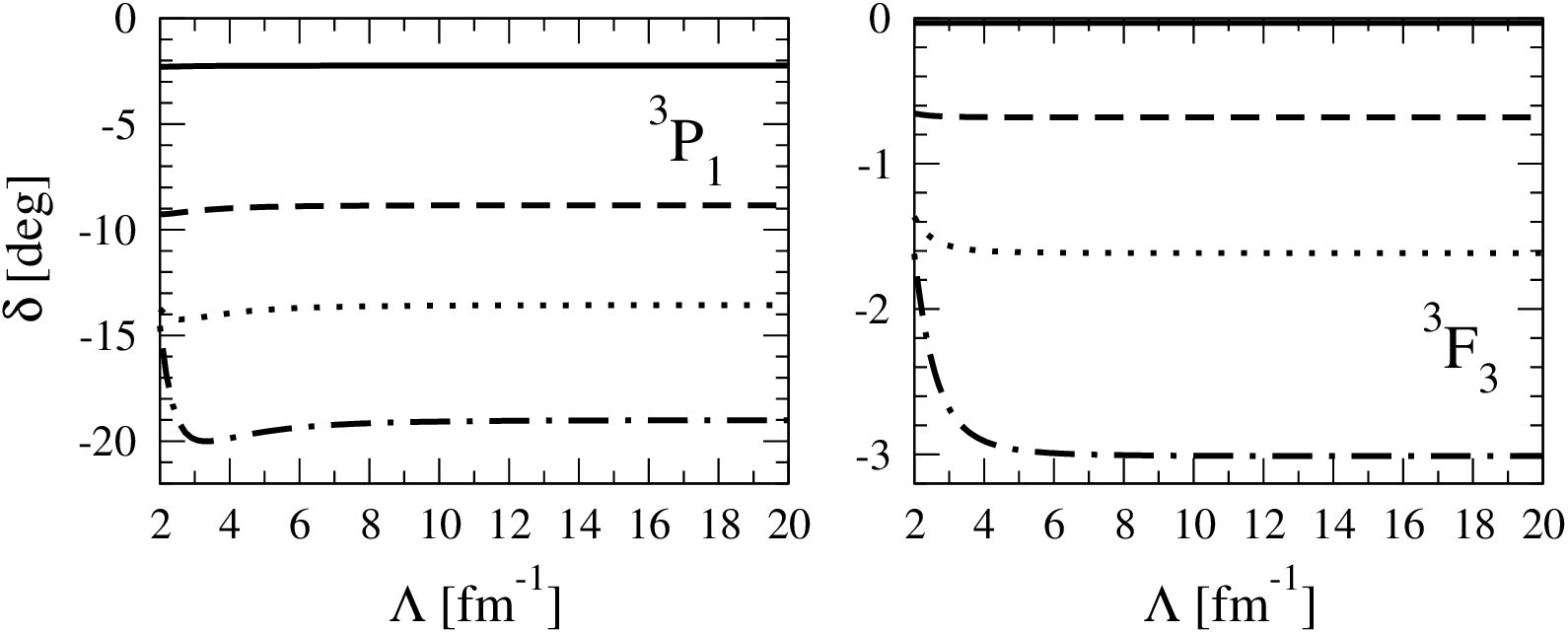}
\end{center}
\vspace{-0.6cm}
\caption{\label{fig:lamdeptrip}  
Cutoff
dependence of the repulsive triplet phase shifts for two partial waves. 
Results are given for laboratory energies of 
10~MeV (solid line), 50~MeV (dashed line), 100~MeV (dotted line), and 190~MeV 
(dash-dotted line).}
\end{figure*}

\begin{figure*}[btp]
\begin{center}
\includegraphics[width=15cm]{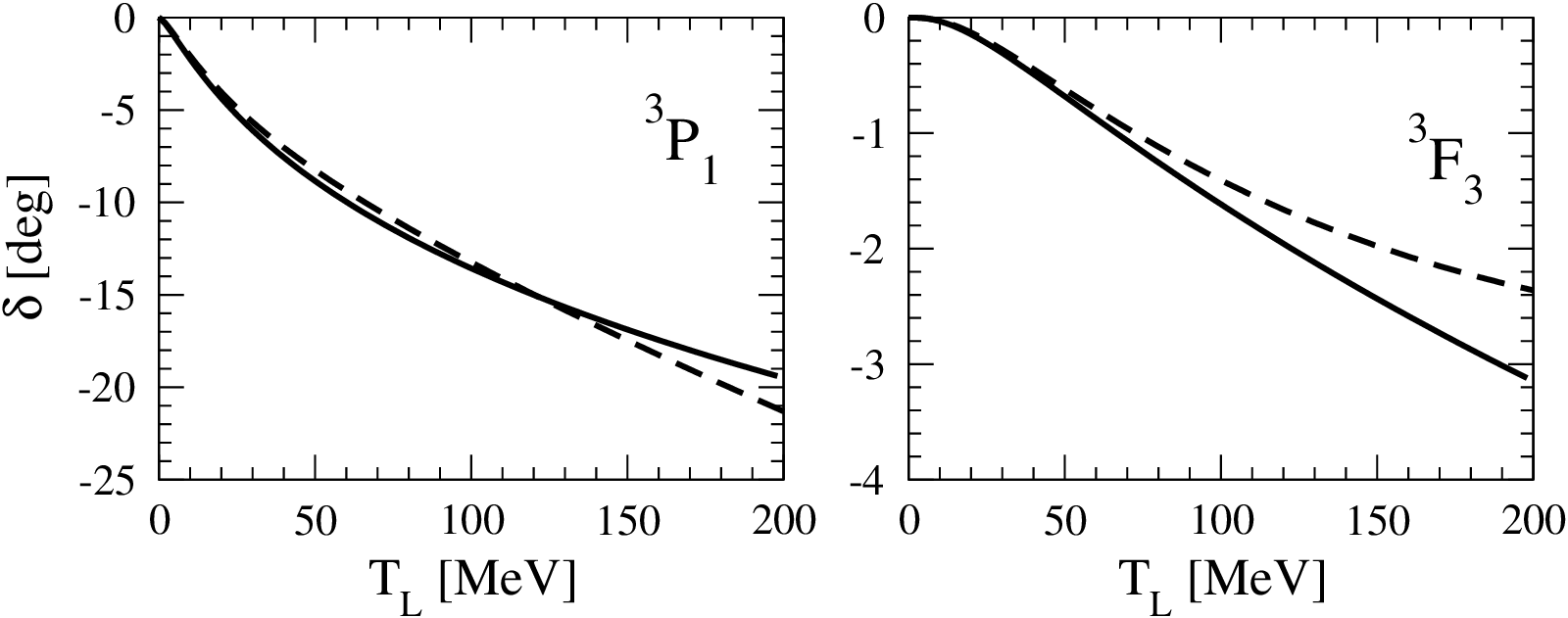}
\end{center}
\vspace{-0.6cm}
\caption{\label{fig:expttrip}  Comparison of LO repulsive triplet 
phase shifts (as function of the laboratory energy)
for $\Lambda=20$~fm$^{-1}$ (solid line)
to the Nijmegen PWA (dashed line).}
\end{figure*}

In the next step, we look at the triplet channels where
the tensor force is repulsive.
The $^3$P$_1$ and $^3$F$_3$ partial waves belong to this class.
The $\Lambda$ dependence of these phase shifts is shown in 
Fig.~\ref{fig:lamdeptrip} and, again, we obtain  $\Lambda$ independent 
results for $\Lambda \simge 5$~fm$^{-1}$ even for energies 
as high as 190~MeV. The comparison with the data
for these cases is shown in Fig.~\ref{fig:expttrip}. 
The $^3$P$_1$ result is in much better agreement with the PWA
than the corresponding result of the $^1$P$_1$ singlet channel. The F wave 
has a comparable accuracy in the triplet and singlet cases.
These results confirm that Weinberg's power counting is, again, consistent 
in channels without an attractive singular interaction. 
The predictions agree well
with the low-energy data in all these cases. 

\begin{figure*}[btp]
\begin{center}
\includegraphics[width=15cm]{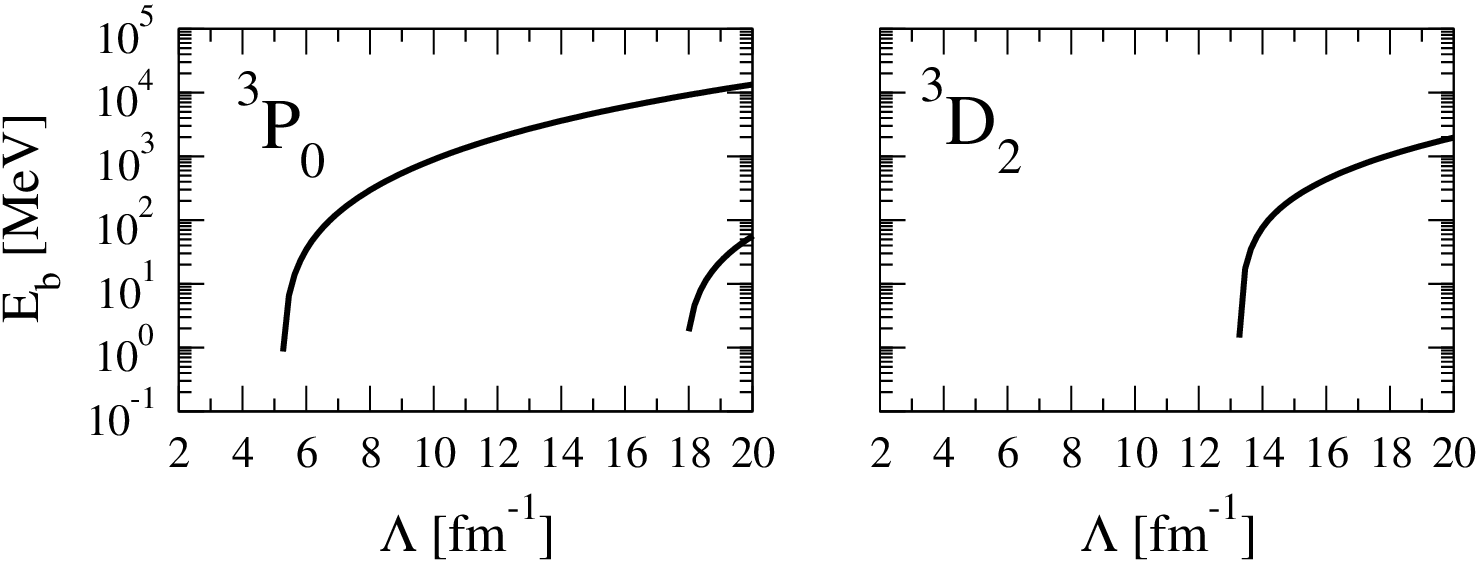}
\end{center}
\vspace{-0.6cm}
\caption{\label{fig:bsnofit}  Binding energies 
of the 
spurious bound states in selected attractive triplet 
channels. }
\end{figure*}

Next, we look at the triplet channels in which an attractive singular
tensor force acts, and therefore
the unregulated problem is not well defined.
At a finite cutoff bound states occur in various waves, and
increasing $\Lambda$ generates more and more of these 
bound states. 
For $\Lambda$ between 2 and 20~fm$^{-1}$,
we find bound states in the $^3$P$_0$ and $^3$D$_2$ channels,
see Fig.~\ref{fig:bsnofit}.
In the 
$^3$P$_2$-$^3$F$_2$ partial waves, we find a bound state 
for cutoffs 
just above $\Lambda=20$~fm$^{-1}$. 
In higher partial waves 
the interaction appears to be screened enough by the centrifugal 
barrier that no bound state occurs in our cutoff range,
although bound states should appear at sufficiently-higher cutoffs.

\begin{figure*}[btp]
\begin{center}
\includegraphics[width=15cm]{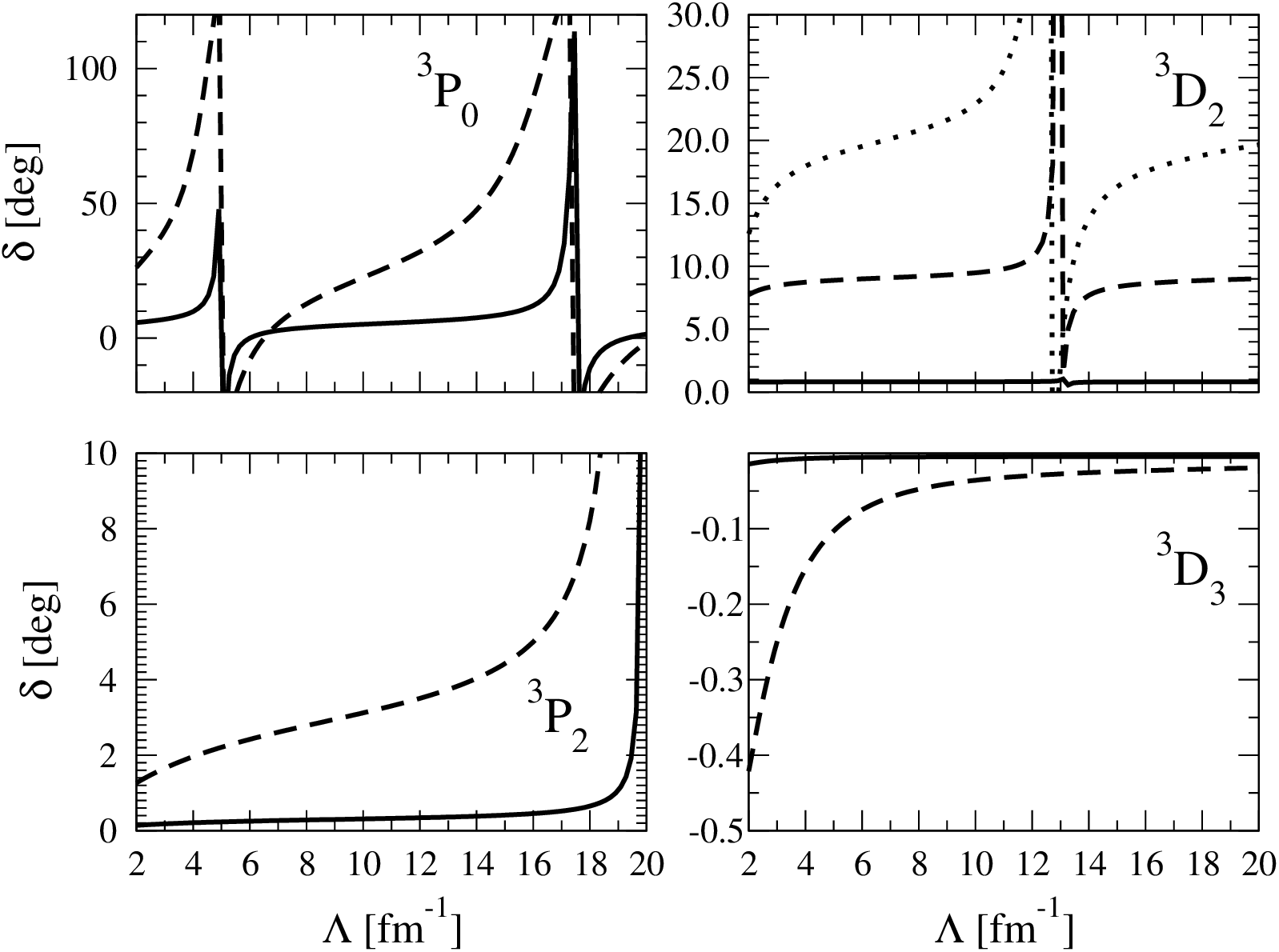}
\end{center}
\vspace{-0.6cm}
\caption{\label{fig:lamdepnofit} 
Cutoff
dependence of phase shifts 
in attractive triplet channels 
at laboratory energies of 10~MeV (solid line), 50~MeV (dashed line),
and 100~MeV (dotted line).} 
\end{figure*}

Even though the binding energies increase rapidly with the cutoff, 
the bound states appear at zero energy. 
It is to be expected that in the cutoff regions 
where new bound states appear the variations of the phase shifts are strong.  
This is explicitly shown in Fig.~\ref{fig:lamdepnofit} 
for the phase shifts at 10~MeV and 50~MeV in various attractive channels.
Clearly, an inconsistency in Weinberg's power counting shows up in these 
channels,
because there are no counterterms available to remove the cutoff dependence
of the observables.

It is interesting to note that in some cases we can
identify cutoff regions in which the results are stable and all the bound
states are deep. 
Some clear plateau regions occur, especially at the lower energies.
At 50~MeV, the $^3$D$_2$ phase shift in the plateau region is 
$\simeq 9^\circ$, which agrees very well with the Nijmegen 
PWA ($8.97^\circ$).
The corresponding $^3$D$_3$ phase shift, however, 
is too small. 
The situation is even worse in the P waves.
At the same energy, 
the cutoff dependence of the $^3$P$_0$ and $^3$P$_2$ phase shifts 
remains visible in any region 
of $\Lambda$s.

This cutoff dependence is related to the singularity of the interaction. 
It is known that an attractive singular central potential requires
a boundary condition 
in each partial wave \cite{behncke68}. 
Therefore, we propose to add to each of the 
problematic triplet channels a counterterm and fit it to experiment, say 
to the phase shift at a certain (low) energy.
We then show that 
the cutoff dependence indeed vanishes also for other energies.
In the following, we will illustrate this
explicitly
for the $^3$P$_0$, $^3$D$_2$, and 
$^3$P$_2$-$^3$F$_2$ partial waves, which are the most problematic 
cases, because bound states exist or are close 
to appearing in the cutoff range that we examined. 
Our results extend the work of 
Refs.~\cite{frederico99,beane02,valderrama04,valderrama05} 
to channels beyond S waves
(and to our choice of regulator). 
  
As argued, we add contact interactions in the $^3$P$_0$ ($i=1$) 
and $^3$P$_2$-$^3$F$_2$ channels ($i=2$) of the form 
\begin{equation}
  V_i = \frac{1}{4} \ \frac{c_i}{(2\pi)^3} \ p' p \ ,  
\end{equation}
which in Weinberg's power counting appear only at 
next-to-leading (NLO) order, or ${\cal O}(Q^2)$. The first D-wave 
counterterms are supposed to be of even higher order: 
N$^3$LO, or 
${\cal O}(Q^4)$. In the $^3$D$_2$ channel, we use  
\begin{equation}
  V_d = \frac{c_d }{(2\pi)^3} \ {p'}^2 p^2 \ . 
\end{equation}

\begin{figure*}[btp]
\begin{center}
\includegraphics[width=15cm]{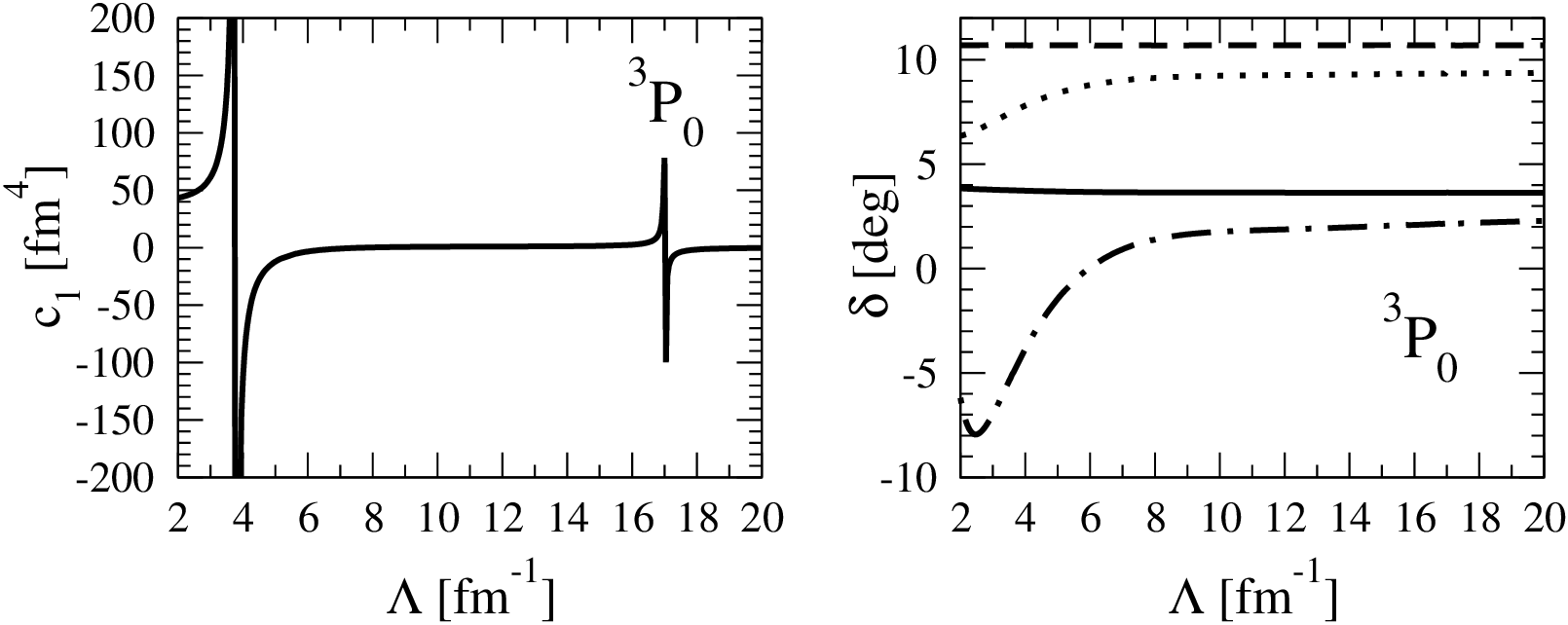}
\end{center}
\vspace{-0.6cm}
\caption{\label{fig:lamdep3p0} Fit result for the counterterm $c_1$ 
as a function of the cutoff,
and the resulting  
cutoff dependence of the $^3$P$_0$ 
phase shift at laboratory energies of 10~MeV 
(solid line), 50~MeV (dashed line), 100~MeV (dotted line),
and 190~MeV (dash-dotted line). }
\end{figure*}

\begin{figure*}[btp]
\begin{center}
\includegraphics[width=15cm]{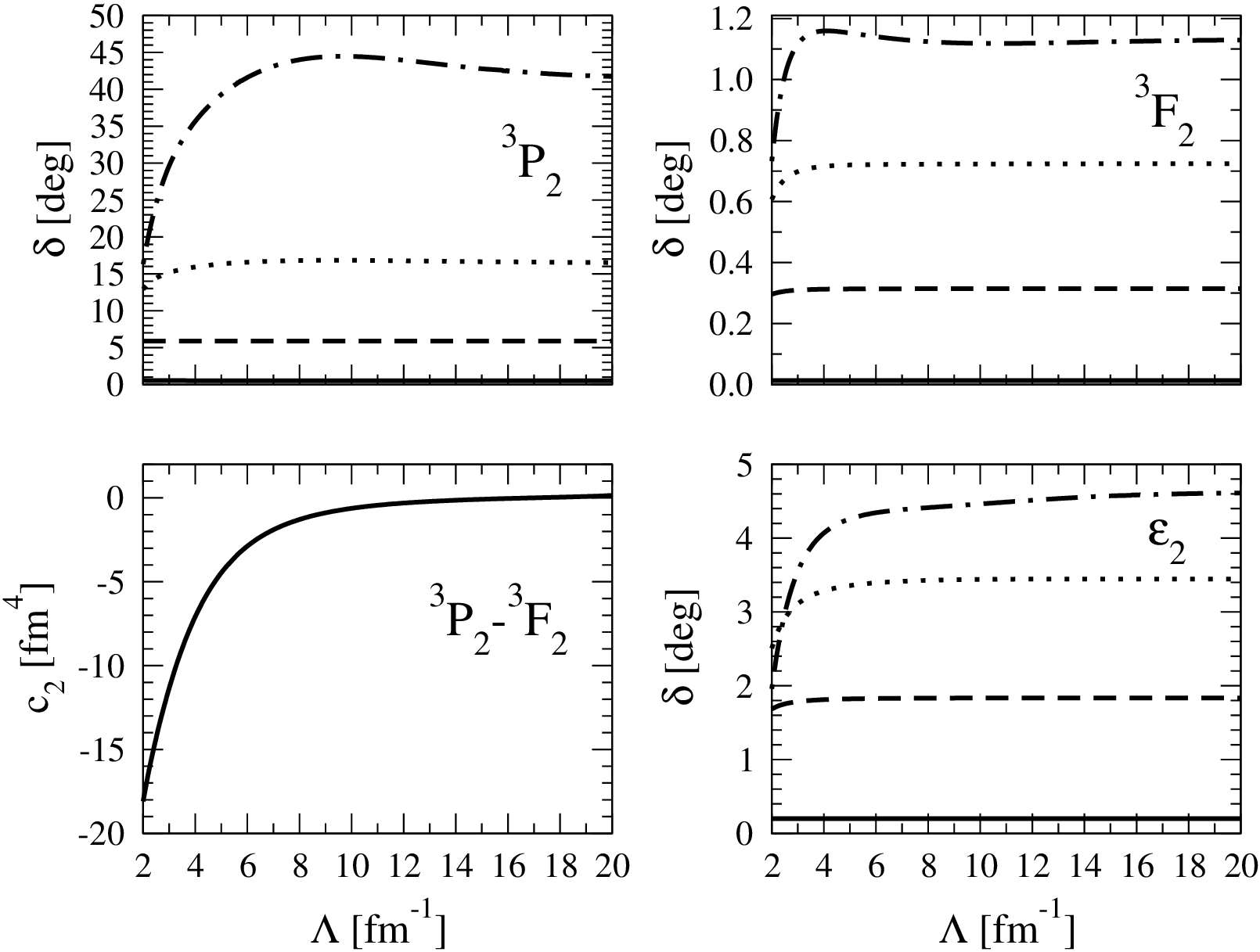}
\end{center}
\vspace{-0.6cm}
\caption{\label{fig:lamdep3p2} Fit result for the counterterm $c_2$ 
as a function of the cutoff,
and the resulting  
cutoff dependence of the 
$^3$P$_2$-$^3$F$_2$ phase shifts
and the mixing angle $\varepsilon_2$
at laboratory energies of 10~MeV (solid line), 50~MeV (dashed line), 
100~MeV (dotted line),
and 190~MeV (dash-dotted line). }
\end{figure*}

\begin{figure*}[btp]
\begin{center}
\includegraphics[width=15cm]{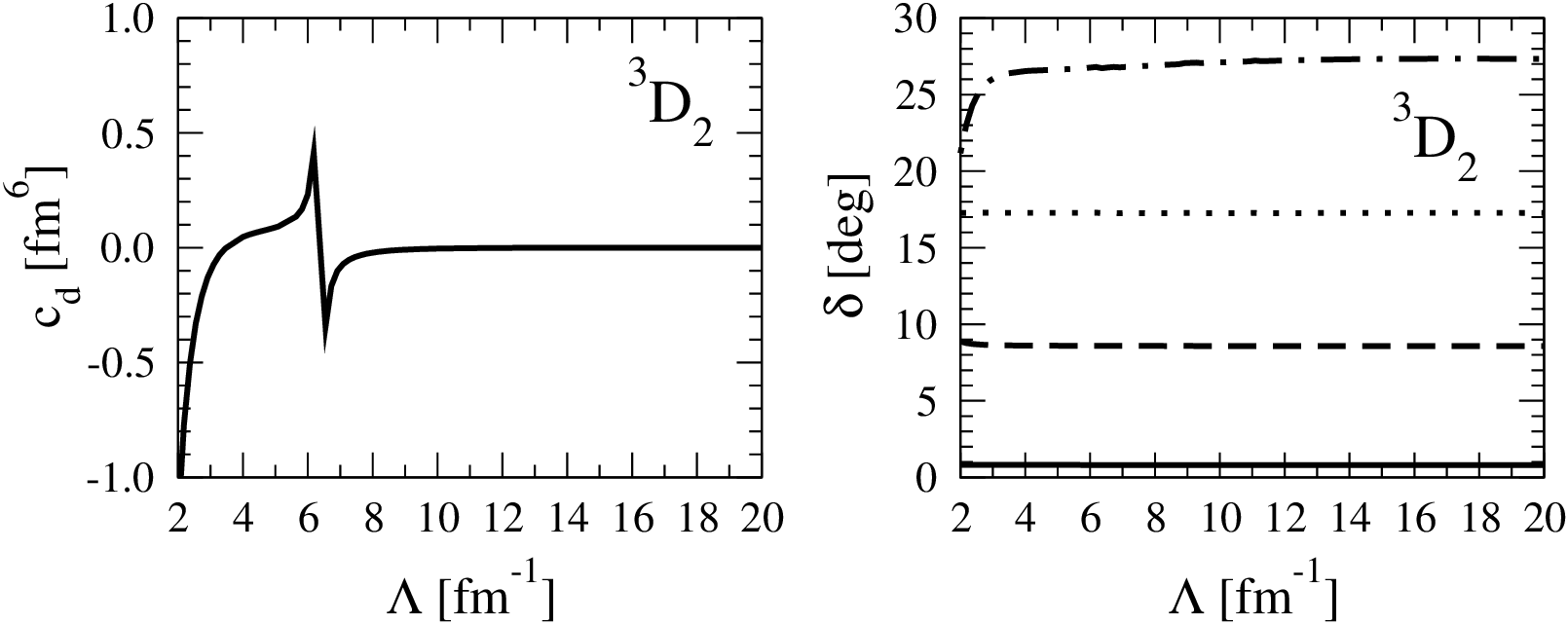}
\end{center}
\vspace{-0.6cm}
\caption{\label{fig:lamdep3d2} Fit result for the counterterm $c_d$ 
as a function of the cutoff,
and the resulting  
cutoff dependence of the $^3$D$_2$ phase
shift at laboratory energies of 10~MeV (solid line), 50~MeV (dashed line), 
100~MeV (dotted line),
and 190~MeV (dash-dotted line). }
\end{figure*}

Fig.~\ref{fig:lamdep3p0} shows our result for the $^3$P$_0$ 
partial wave. 
The value of $c_1$ was determined by a fit of the phase shift 
for a laboratory energy of 50~MeV. Since the size of the counterterm 
is not bounded, we varied 
this constant by orders of magnitude, but could not find any further 
solution that describes the phase shifts equally well.
The cutoff dependence of $c_1$  exhibits a nice limit-cycle-like behavior,
similar to that of $c_t$. 
Fig.~\ref{fig:lamdep3p0} also demonstrates that the resulting phase shifts 
at other energies are cutoff independent for $\Lambda \simge 8$~fm$^{-1}$. 
Figs.~\ref{fig:lamdep3p2}  and \ref{fig:lamdep3d2} summarize the 
analogous results for the $^3$P$_2$-$^3$F$_2$ and  $^3$D$_2$ partial
waves, respectively. 
The fits were performed using the $^3$P$_2$ phase shift at  50~MeV and 
the  $^3$D$_2$ phase shift at 100~MeV. We confirm the cutoff independence 
(for large $\Lambda$) in all phase shifts and mixing parameters. 

An alternative 
to absorbing the cutoff dependence in the various P waves individually
would be to employ 
one counterterm with tensor structure. 
Unfortunately, we have not been able to implement this idea
without introducing 
cutoff dependence in the $^3$P$_1$ wave. 

\begin{figure}[btp]
\begin{center}
\includegraphics[width=6cm]{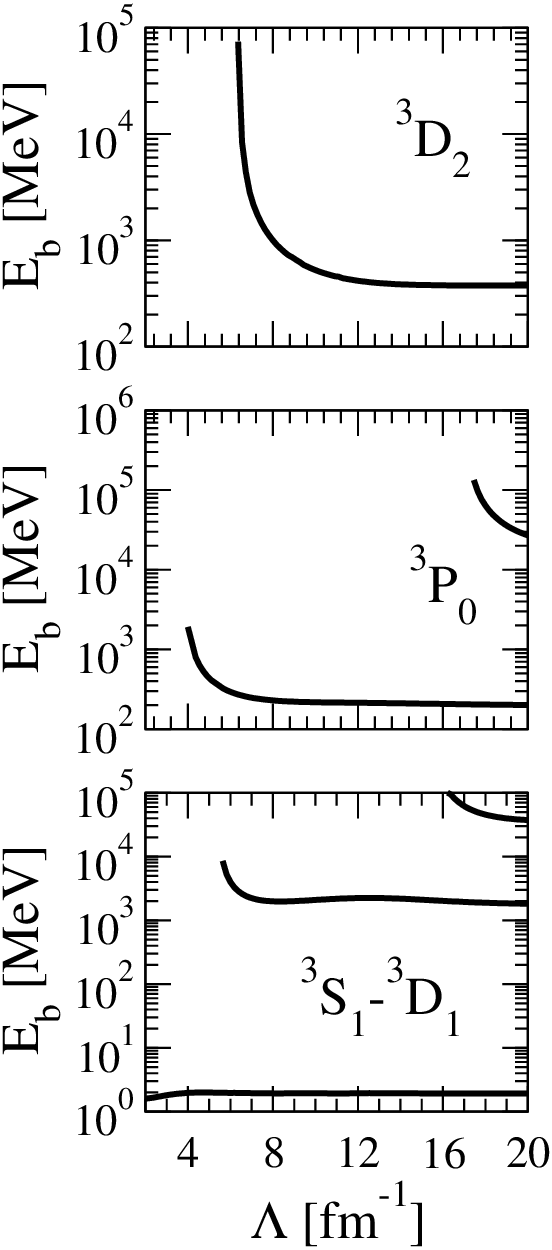}
\end{center}
\vspace{-0.6cm}
\caption{\label{fig:boundfit} Binding energies of bound states
found in various partial waves as function of the cutoff.
The shallow bound state in the $^3$S$_1$-$^3$D$_1$ coupled channels
corresponds to the deuteron, while the other, deep, bound states are
outside the range of applicability of the EFT.}
\end{figure}

After removing the cutoff dependence by adding 
appropriate counterterms, we still find 
spurious bound states in the $^3$P$_0$, $^3$D$_2$, and also the
$^3$S$_1$-$^3$D$_1$ channels. However, the cutoff dependence of the
binding energies is now
completely different, as shown in Fig.~\ref{fig:boundfit}.
As desired, only $^3$S$_1$-$^3$D$_1$ has a shallow 
bound state, the deuteron, which
is cutoff independent over almost the entire 
$\Lambda$ range; the deuteron binding energy is predicted to be
$1.92$~MeV in this LO calculation. 
The bound states in the other channels are all very deep. A new bound
state appears with infinite binding energy around the cutoff at which 
the corresponding counterterm is singular,
and then approaches a constant, large binding energy for
increasing $\Lambda$. 
These bound states are beyond the range of the EFT,
and they are irrelevant for the low-energy physics.

\begin{figure*}[btp]
\begin{center}
\includegraphics[width=15cm]{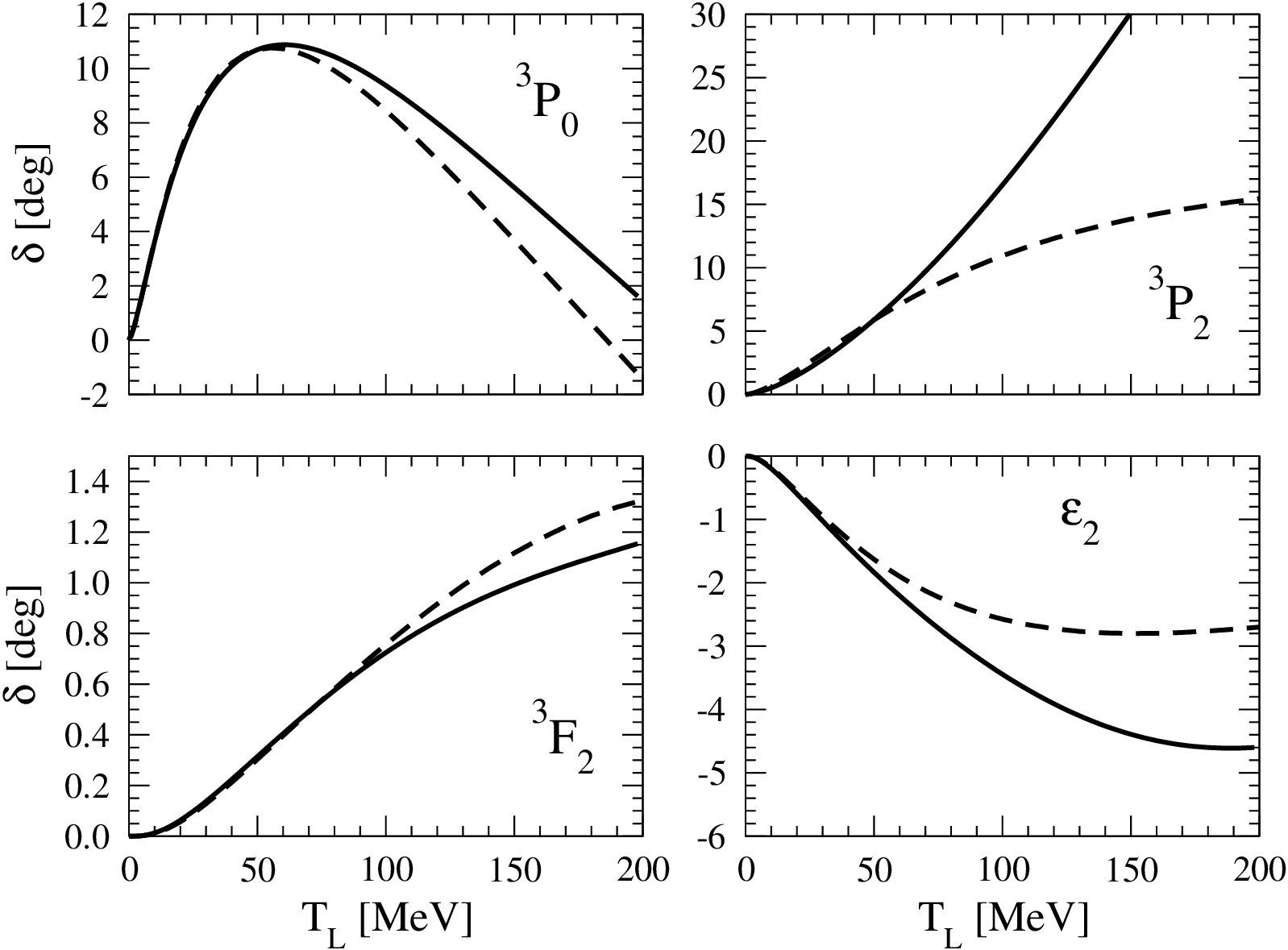}
\end{center}
\vspace{-0.6cm}
\caption{\label{fig:phasepwave} Comparison of 
attractive triplet phase shifts 
(as function of the laboratory energy)
for  $\Lambda=20$~fm$^{-1}$ (solid line)
to the Nijmegen PWA (dashed line).}
\end{figure*}

With the added counterterms,
we obtain a very decent description 
of the phase shifts. 
Fig.~\ref{fig:phasepwave} shows that our 
$^3$P$_0$ 
result follows the energy-dependence of the Nijmegen PWA
remarkably well. Obviously, the addition of the counterterm is here
supported by the experimental data. 
In the coupled $^3$P$_2$-$^3$F$_2$
channels the agreement with the PWA below $50$~MeV is still satisfactory. 
We emphasize that the $^3$F$_2$ phase and the mixing parameter 
$\varepsilon_2$
are predictions. Choosing a high cutoff $\Lambda$ clearly does not compromise 
the description of these observables.  

\begin{figure*}[btp]
\begin{center}
\includegraphics[width=15cm]{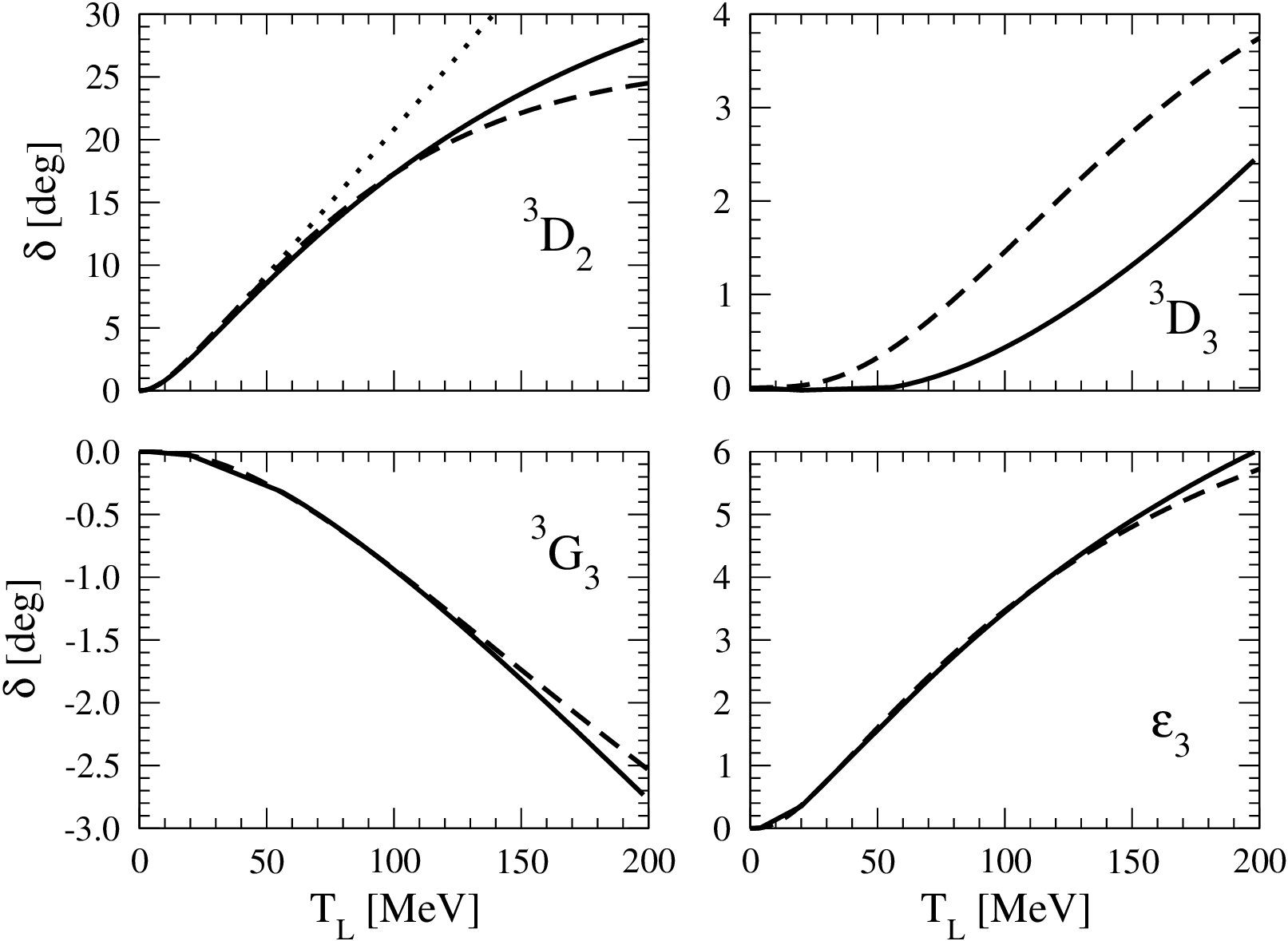}
\end{center}
\vspace{-0.6cm}
\caption{\label{fig:phasedwave} Comparison of attractive triplet phase shifts
(as function of the laboratory energy)
for  $\Lambda=20$~fm$^{-1}$ (solid line)
to the Nijmegen PWA (dashed line). 
For $^3$D$_2$ also 
the result for $\Lambda=8.0$~fm$^{-1}$ without counterterm is 
given (dotted line).}
\end{figure*}

For the $^3$D$_2$ phase, see Fig.~\ref{fig:phasedwave}, we find again 
a good agreement with the PWA. Here, we also included the prediction 
based on a calculation without counterterm, for $\Lambda=8.0$~fm$^{-1}$ 
in the plateau region of Fig.~\ref{fig:lamdepnofit}. For low
energies below 50~MeV, the results are comparable. The deviations from
the PWA become significant toward higher energies, where the 
plateau seen in Fig.~\ref{fig:lamdepnofit} is more and more tilted. 
For these higher energies, the  counterterm again improves the predictions.  

\begin{figure*}[btp]
\begin{center}
\includegraphics[width=15cm]{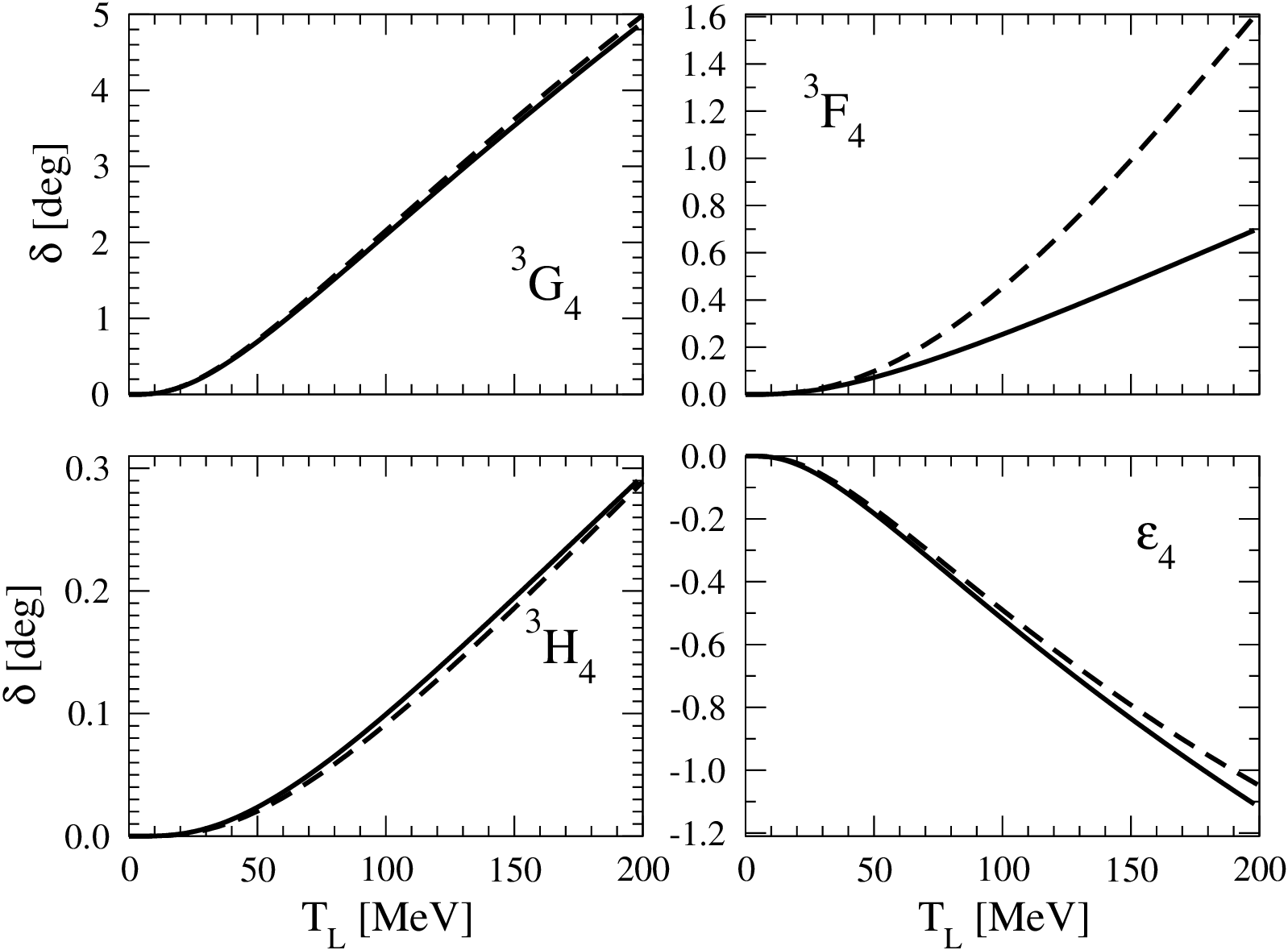}
\end{center}
\vspace{-0.6cm}
\caption{\label{fig:phasefwave} Comparison of attractive triplet phase shifts
(as function of the laboratory energy) 
for  $\Lambda=20$~fm$^{-1}$ (solid line)
to the Nijmegen PWA (dashed line).}
\vspace{0.6cm}
\end{figure*}

Our overview is completed in Figs.~\ref{fig:phasedwave} 
and~\ref{fig:phasefwave} with the $^3$D$_3$-$^3$G$_3$, $^3$F$_4$-$^3$H$_4$,
and $^3$G$_4$ channels. 
In these partial waves there is a relatively small cutoff dependence
in the $\Lambda$ range that we studied
(although presumably cutoff dependence will become significant
at cutoffs high enough to bring in spurious bound states).
In all cases the agreement with the PWA is improved when we increase
the cutoff from the traditional values around 2.5~fm$^{-1}$ \cite{epelbaum00} 
to our higher values.
This is especially true for the $^3$D$_3$ partial wave, which, for 
our higher cutoffs, becomes attractive for higher energies. 

After these encouraging results, we examine the $3N$ bound state
in the next section.

\section{Three-nucleon bound state}
\label{sec:tri}

The power of EFT comes to bear when more nucleons are considered.
The $3N$ system
is the first extension to few-nucleon systems
to consider. 
According to Weinberg's power counting, $3N$ forces
should be subleading in the pion\-ful theory, even though there
is a leading-order $3N$ force in the pionless EFT \cite{bedaque00}.
Calculations 
of the triton properties using traditionally-low values for $\Lambda$ 
have been published in Refs. \cite{epelbaum01a,epelbaum02a,epelbaum02c}
for NLO and N$^2$LO in Weinberg's power counting,
with and without $3N$ interactions. 
So far, a LO calculation was omitted, because of the unsatisfactory
description of the $N\!N$ phase shifts, especially $^1$S$_0$. 
Here, our main goal is to assess the ordering of $3N$ counterterms. 
For this 
purpose we do not require a high-quality description of the $N\!N$ 
phase shifts, but instead we need to study the dependence of 
the triton binding energy on 
a larger range of cutoffs.

The $^1$S$_0$ problem cannot be addressed in this work, but a calculation 
of the triton binding energy ($E_t$) is of interest in order to 
compare the renormalization of the $3N$ system
in the pion\-ful theory with that in the pionless EFT.
We will demonstrate
that no $3N$ counterterms are necessary to ensure cutoff 
independence in LO in the pion\-ful theory,
once the LO calculation has been modified according to the previous section. 
This result will be important in future studies of 
the convergence of the chiral expansion 
in few-nucleon systems. The appearance of $3N$ interactions 
at N$^2$LO, the parameters of which are generally adjusted to the experimental
value of $E_t$,
makes predictions in LO and NLO especially important in this respect.  
We defer a study of subleading orders and $N\!d$ scattering states to 
a later publication.

We calculate $E_t$ by solving the Faddeev equations for the $3N$ system.
The techniques 
were recently described, {\it e.g.}, in Ref. \cite{nogga02b}.  Here we have
to deal with the additional complication that the $N\!N$ interaction supports
deep, spurious bound states in various partial waves, which we remove as
described in Appendix~\ref{app:tmat}. 
We confirmed the accuracy of this prescription by comparing our results
for the energy to the expectation value of the Hamiltonian using the unaltered 
interaction. Both values agreed within several keV.

\begin{figure}[btp]
\begin{center}
\includegraphics[width=8.5cm]{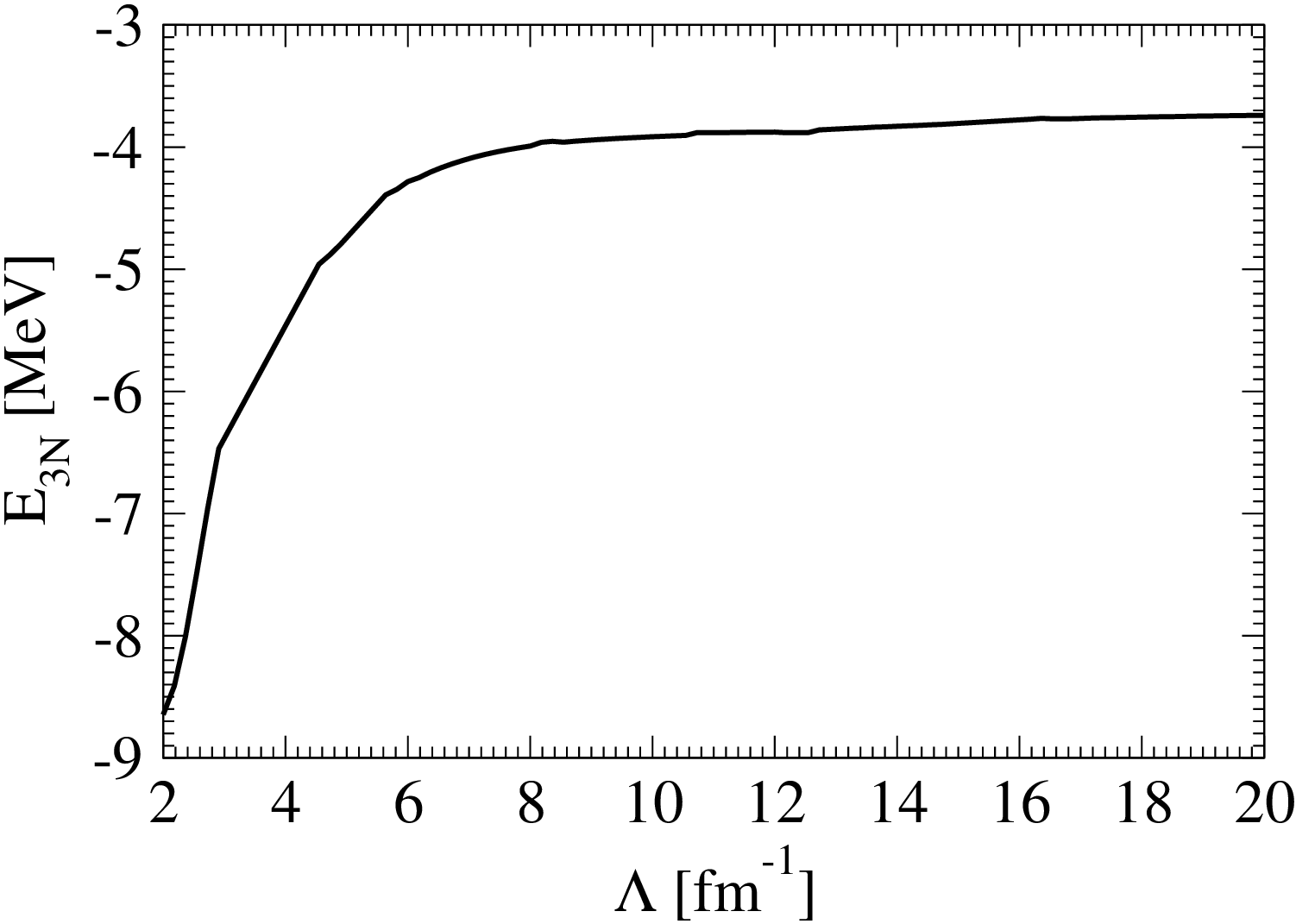}
\end{center}
\vspace{-0.6cm}
\caption{\label{fig:3nbind} 
Cutoff dependence of the triton binding energy.}
\end{figure}

\begin{figure}[btp]
\begin{center}
\includegraphics[width=8.5cm]{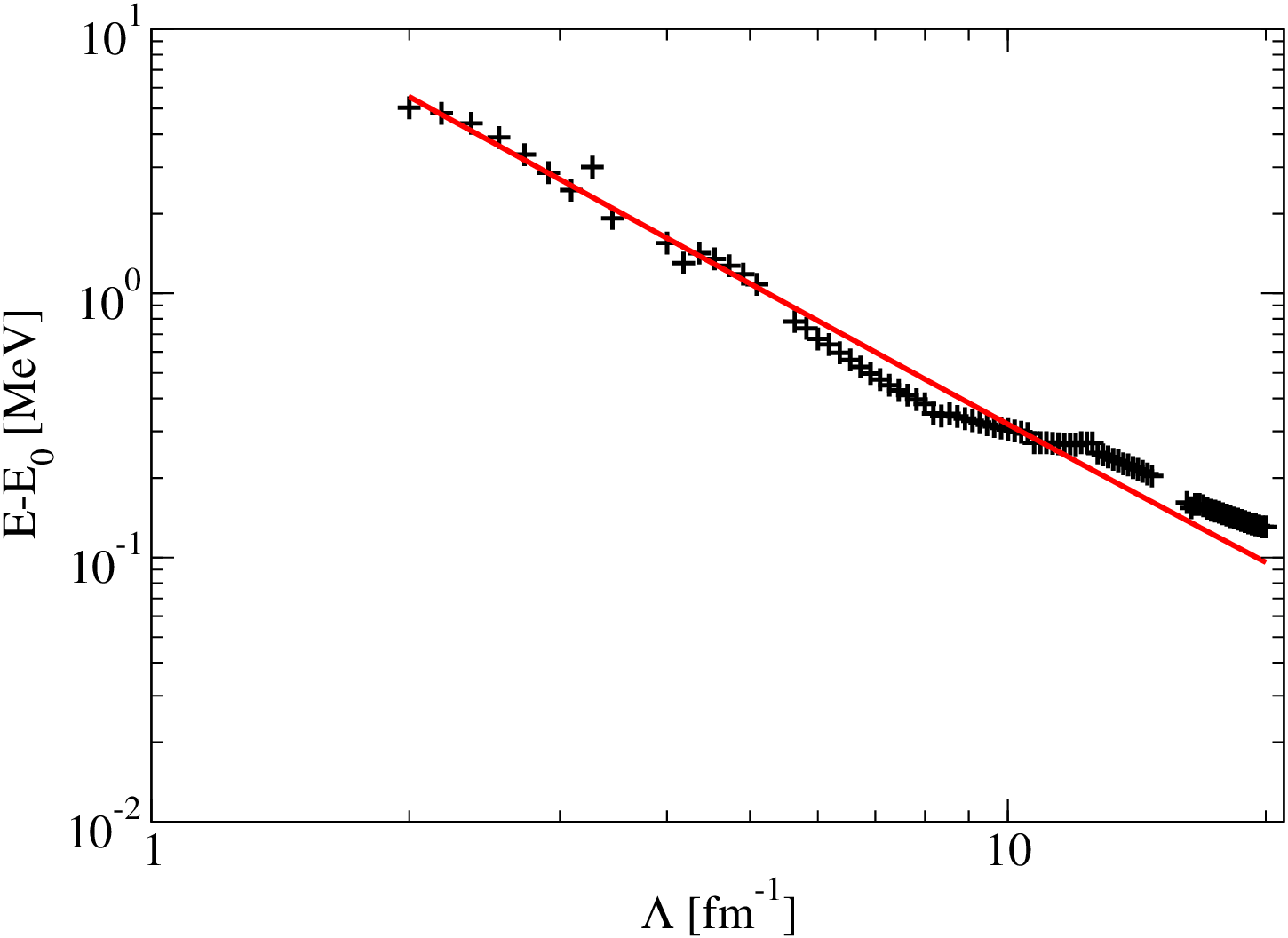}
\end{center}
\vspace{-0.6cm}
\caption{\label{fig:3nlepage} 
Numerical results (crosses) and fit (solid line)
for the deviation of the triton binding energy from
its converged value, as function of the cutoff.
See text for details.} 
\end{figure}

The cutoff dependence of $E_t$ is shown in Fig.~\ref{fig:3nbind}.
We see a plateau region starting around $\Lambda = 8$~fm$^{-1}$.
To extract the converged result from this calculation and to 
confirm that only terms of order ${\cal O}(Q^2/\Lambda^2)$ 
and higher are missing, 
we fitted the function
$E(x) = E_0 \ (1+(C / \Lambda)^x)$ to our numerical results. We obtained
the converged binding energy $E_0=-3.6$~MeV, $C=2.54$~fm$^{-1}$, and 
$x\simeq 1.8$.  
The exponent is in reasonable agreement with the expectation $x=2$.
The quality of this approximation to the cutoff dependence can be
observed in the Lepage plot \cite{lepage97} of Fig. \ref{fig:3nlepage}, where
we show $E-E_0$ versus $\Lambda$ 
on a double-logarithmic scale. The data follows the fitted results nicely.   
For high values of $\Lambda$ the slope seems to change slightly. 
This can be a numerical 
artifact, because $E-E_0$ is rather small in this range, which increases
the relative, numerical uncertainty.

A cutoff dependence as it occurs in the $3N$ system within the 
pionless EFT \cite{bedaque00}
does not appear here. 
We stress that, in contrast, for 
the same range of $\Lambda$ we found considerable variations 
in  several $N\!N$ phase shifts before adding counterterms. 
This gives us confidence 
that our cutoff range is large enough to draw
conclusions about the order of the counterterms.
In particular,
we conclude that the finite range of 
OPE prevents the Thomas collapse of the $3N$ bound state \cite{thomas35}. 

Our LO results imply a rather-large underbinding of the triton in the limit 
of high $\Lambda$. In the region around $\Lambda=2.5$~fm$^{-1}$, 
our prediction is in much better agreement with the experimental 
value $E_t=-8.48$~MeV. For those cutoffs the Weinberg NLO predictions 
are very similar \cite{epelbaum01a}. 
This suggests that the NLO result might be 
less cutoff dependent than the LO result, and that 
the plateau region starts at lower values of $\Lambda$ 
for higher orders. Our result is probably influenced significantly by the 
unrealistic description of the $^1$S$_0$ $N\!N$ phase shift,
which will improve in a NLO calculation. 
Binding energies are more sensitive to higher orders, 
because in a theory with pions
the potential energy is to a large extent canceled by the kinetic 
energy. 
In view of this, we are not very concerned about  
our LO plateau value.
It will be interesting to analyze the NLO results in  a similar way. 
If our expectations are
verified, they would reconcile our observations with the good 
results obtained with a fixed, low cutoff
in Refs. \cite{entem03a,epelbaum02c}.

\section{Lessons for Power Counting}
\label{sec:counting}

It has been realized for some time now that
EFT power counting is more
complicated for non-perturbative than
for perturbative processes.
In particular, one has to consider the effects
of infrared enhancements in the running of counterterms,
which can invalidate naive dimensional analysis. In this section,
we discuss some of the implications of our findings to power counting
in the pion\-ful EFT.

The existence of shallow (real and virtual) bound states in both $N\!N$
S waves is a clear sign that non-perturbative physics is taking
place, in contrast to the situation in ordinary ChPT.
We can describe this in the same language used to
discuss power counting in ChPT \cite{weinberg79,weinberg90,weinberg91}:
we represent typical nucleon momenta by $Q$
and the characteristic scale of QCD in the hadronic phase by $M_{QC\!D}$.
The effect of iterating an interaction in the kernel of the $T$ matrix 
is twofold. First, one has 
an extra three-dimensional momentum integral and an
extra $N\!N$ Schr\"odinger propagator.
Second, one has an extra factor of the potential.
After the cutoff dependence is removed by renormalization,
the contribution to the $N\!N$ $T$ matrix from
an $N\!N$ intermediate state is expected to be ${\cal O}(m_N Q/4\pi)$. 
This is a factor $m_N/Q \gg 1$ 
larger than in analogous states in ordinary ChPT,
and it is due to the small energy of intermediate states 
containing nucleons only.
If the interaction has an effective strength $v_{l'l}$ connecting waves
of orbital angular momenta $l$ and $l'$,
one iteration then roughly brings a dimensionless 
factor ${\cal O}(m_N Q v_{l'l}/4\pi)$.
For $Q\simge 4\pi/m_N v_{l'l}$ 
this factor is $\simge 1$ and the
interaction has to be iterated to all orders, potentially
leading to bound states.

The pion\-ful theory is relevant for momenta $Q\simge m_\pi$. 
In this case, we estimate the effect of OPE 
as $v_{l'l} \sim \alpha_{l'l}/f_\pi^2$,
where $\alpha_{l'l}$ is a dimensionless angular-momentum factor.
Therefore, we can expect OPE to be non-perturbative 
at $Q\simge 4\pi f_\pi^2/m_N \alpha_{l'l}$.
Ignoring the angular-momentum factor,
we get $Q\simge 100$ MeV for the momentum
where pions need to be iterated.
This is in agreement with what is observed in a perturbative
calculation of low waves, where
twice-iterated pion exchange seems to overcome 
one-pion exchange in
various channels for momenta around 100 MeV \cite{fleming00}. 

In spin-singlet channels, OPE goes as $1 -m_\pi^2/q^2 +\ldots$
at high momentum. 
When iterated,
the first term by itself introduces cutoff dependence in the S wave
only, which can be removed by a chirally-symmetric
counterterm $C_0$.
The interference between the iteration of this counterterm and
the second term in OPE
generates further cutoff dependence in the $^1$S$_0$ wave,
which in turn can be removed \cite{kaplan96,beane02} 
by a chiral-breaking counterterm
$m_\pi^2 D_2$, {\em cf.} Eq.~(\ref{C0D2}).
This counterterm is enhanced with respect to naive dimensional
analysis, and should be promoted to LO if pion-mass effects
are kept in LO, as it seems most efficient.
In this paper we have checked explicitly that 
OPE is well behaved in the higher spin-singlet waves. 
(Higher spin-singlet waves have also been recently discussed
in Ref. \cite{birse04}.)

In spin-triplet channels, the situation is complicated 
by the tensor operator, which retains angular dependence
even asymptotically. 
As shown in Refs. \cite{frederico99,beane02} and confirmed here, 
the cutoff dependence introduced by iteration in the $^3$S$_1$-$^3$D$_1$
coupled channel can be dealt with a single
chirally-symmetric, momentum-independent counterterm,
as prescribed by Weinberg's power counting.
The deuteron and $^1$S$_0$ virtual state have,
for the observed value of the quark masses, binding
momenta somewhat smaller than our crude estimate 
$Q\sim 4\pi f_\pi^2/m_N$, indicating an amount of fine tuning.
However,
if one varies the pion mass the momentum scales 
for the bound states acquire more natural values 
\cite{beane02,beane03b,epelbaum03d,beane03c}.

Yet, as we have shown above, iterated OPE produces 
spurious bound states and 
cutoff
dependence in all waves where the tensor force is attractive.
This undesired feature can only be removed by 
additional counterterms
in the corresponding waves. 
Since Weinberg's power counting only prescribes
counterterms in the S waves, our results clearly 
upset Weinberg's power counting.

Weinberg's power counting was based on naive dimensional analysis.
Contact interactions are necessary to
remove divergences from loops that do not involve purely-nucleonic
intermediate states.
These loops are not infrared enhanced,
and it is reasonable to assume that they scale ---as in
ordinary ChPT--- with powers of $(Q/4\pi f_\pi)^2$.
This has in fact been confirmed by explicit calculation 
\cite{ordonez92,kaiser97,kaiser98}. 
Implicit in Weinberg's power counting is the assumption 
that loops in the LS equation
do not bring significant new cutoff dependence.
The parameters of
contact interactions with derivatives or powers of the
pion mass would thus be suppressed
by powers of a large mass scale, $M_{QC\!D}$, and
the effects of derivatives would scale as $Q/M_{QC\!D}$.

However, we now see that Weinberg's implicit assumption is 
not correct.
The short-range parameters needed to renormalize iterated OPE
do not obey naive dimensional analysis because their
renormalization-group running is enhanced in the infrared.
These counterterms are driven by pion parameters,
and the effects of derivatives can scale as $Q/f_\pi$.
(If we take $m_N\sim 4\pi f_\pi$, there is no other
dimensionful parameter at LO than $f_\pi$.)
Taken at face value, this implies that all these counterterms
must be considered in
leading order. 
This is not a complete disaster, as there is still some predictive 
power left, for example in the energy dependence of each 
attractive partial wave, and in the repulsive waves.
However, it would put few-nucleon observables that include significant
contributions from many partial waves, such as the triton binding energy,
out of reach.

In the following, we want to motivate that this complication 
can be avoided in higher partial waves, because these 
are still perturbative.
We therefore consider $\alpha_{l'l}$,
which involves a kinematic
suppression that accounts for the repulsive
effect of the centrifugal barrier. 
The appropriate counterterms will make OPE well defined,
by selecting in the region $r\sim 1/f_\pi$ the correct
combination of solutions for the long-range potential \cite{beane01,beane02}. 
Therefore, the kinematic suppression can be estimated 
as for a regular potential. 
In the case of
a central potential, it can be shown \cite{goldberger75}
that for $l\gg Qd$, where $d$ is the range of the interaction,
the $l$-wave phase shift is given by
$\tan \delta_l\sim (Qd/(l+1/2))^{2l+1} \ll 1$
(barring the exceptional case of a fine tunning that generates 
a $T$-matrix pole near the origin of the complex momentum plane).
This is consistent with the expectation that in the 
classical limit there is little scattering
when the impact parameter $l/Q$ is much larger than the range $d$.
The ratio of the $T$ matrix, and thus the potential, between $l+1$ and $l$
is ${\cal O} (Q/lm_\pi)^2$, for large $l$. For $Q\sim m_\pi$,
we are led to $\alpha_{l,l}= {\cal O} (1/l!^2)$.
In the case of the tensor force, 
there must be two different factors $1/l!$
and $1/l'!$,
in addition to the elements in Table \ref{tab:tensor},
which do depend on $l$ and $l'$, but approach 
constant values for large $l$, $l'$.
Although at small $l$ and $l'$ $\alpha_{l'l}$ can be very complicated,
we expect that 
$\alpha_{l'l}={\cal O}(1/l'! l!)$
for large $l$, $l'$. For this argumentation we assumed that the 
pion mass is finite. We note that it does not apply in the chiral limit. 

This qualitative argument suggests that the effects of the corresponding
higher-derivative counterterms
are suppressed by a large (for large $l$) scale $l f_\pi$.
Obviously, there might be other dimensionless factors that we
miss here, but the fact that factors of $l$ suppress
OPE and its required counterterms in high-$l$ waves must hold.
There are several implications of this new counting
that seem to be supported by existing results.

The kinematic suppression of
higher waves makes the strength of OPE weaker with increasing
$l$. 
In high waves, OPE and probably all  subleading interactions can 
be treated in perturbation theory even for momenta of the order 
of $m_\pi$. We therefore find explicitly that high partial 
waves can be treated perturbatively. 
The perturbative nature of OPE
in high waves is part of nuclear folklore, and has been checked in 
EFT explicitly \cite{kaiser97,kaiser98,epelbaum04a}.

For sufficiently-large $l$, the suppression factor in counterterms 
becomes dominated by  $M_{QC\!D}$ (rather than $lf_\pi$), representing 
omitted QCD degrees of freedom,
and the size of the counterterms is that assumed in
Weinberg's power counting.

On the other hand,
for a finite number of low partial waves, {\it e.g.} $^3$P$_0$ and  
$^3$P$_2$-$^3$F$_2$, we find that perturbation theory is not 
sufficient for momenta of the order of $m_\pi$. This is caused 
by the lack of enough suppression from $l$,
and by unnaturally-large $|\alpha_{ll'}|$ in these cases. 
Resummation is necessary and needs to be performed 
numerically. 
Our numerical analysis showed that the cutoff dependence can be absorbed 
by one counterterm per partial wave. 
The favorable agreement of our LO calculation
with the data indicates that we did not introduce additional 
inconsistencies and that the resummation includes the most important diagrams.

We conjecture that this mixture of perturbative treatment of higher 
partial waves, resummation of lower partial waves, and promotion 
of a finite number of counterterms is the most consistent approach 
to ChPT for nuclear systems. In subleading orders, it would naturally suggest 
a perturbative treatment of the subleading interactions, 
if no unnaturally-large 
$\alpha_{ll'}$s are present.
The NLO interactions consist in principle
of TPE and counterterms with two more derivatives
than LO.
Subsequent orders are constructed by the inclusion
of successive powers of $Q/M_{QC\!D}$.
The most-effective organizational scheme 
for subleading interactions probably relies
on taking into account an explicit delta-isobar field 
\cite{ordonez94,kolck94,ordonez96,kaiser98,pandharipande05}.
The correctness of our modified power counting needs, of course, to be 
checked in future studies of higher orders.  

On the other hand, for practical reasons, it might be convenient to  
perform the resummation in all partial waves that are taken 
into account. Iterating something small causes only a small error, so
one might decide to iterate OPE in all waves, as done automatically when
solving the LS for OPE. This is again part of nuclear
folklore. If we do this without the corresponding counterterms,
however, cutoff dependence is introduced. By increasing
the cutoff, the iteration of OPE can be made arbitrarily large,
and at some point bound states appear. 
The kinematic suppression suggests
that the cutoffs for which bound states appear increase
with $l$, which is consistent with what we observe
in the cutoff window we studied.
Existing calculations based on Weinberg's
counting 
should exhibit \cite{meissner01a} some of the problems we point out here.
In particular, as the cutoff is increased, partial waves
without the required counterterms become unwieldy. 

In that case, one should consider cutoffs
in a limited range, for LO between, say, 5 to 10 fm$^{-1}$.
Variation of the cutoff in this limited range
would not exhibit any of the drawbacks pointed out here,
because it leads to an effective suppression of the higher-order terms. 
Therefore, reasonable fits to the data can be achieved.
This explains the success of existing fits 
\cite{epelbaum00,entem02b,entem02a,entem03a,epelbaum05a}
over a limited cutoff range. 

The variation of the cutoff within a given range
will generate a band of values for observables.
The error in a fit based on 
Weinberg's counting
is likely dominated by the lowest partial wave without the
required counterterm. As one goes to higher orders in Weinberg's counting,
one acquires more counterterms, pushing
the error to higher waves.
The $l$ suppression then ensures that the bands for the observables
shrink, as observed \cite{epelbaum00,entem02b,entem02a,entem03a,epelbaum05a}.
This does {\em not}, however,
imply that Weinberg's power counting is correct.

At any given order, our modified power counting 
has more short-range parameters than the same order in 
Weinberg's power counting.
In this context, it is interesting 
to note that existing results in Weinberg's power counting
already suggest that short-range parameters
are relatively more important than the long-range physics
that is supposed to be of the same order.
For example,
Refs. \cite{entem02b,entem02a} considered N$^2$LO in Weinberg's power
counting plus D-wave counterterms and found good fits.
These calculations have a couple of counterterms more than what we advocate
here, but they come pretty close to our N$^2$LO.

A similar observation can be made about the Nijmegen PWA when we look at
it from the point of view of chiral EFT, with the PWA short-range parameters
playing the role of counterterms. The long-range strong interaction
consists of OPE and (leading and subleading) TPE~\cite{rentmeester99,
rentmeester03}, and thus corresponds to
N$^2$LO in Weinberg's power counting. In the Nijmegen PWA, short-range
parameters are added for the various partial waves until the fit to the
data, up to 350 MeV laboratory energy, is optimal. (The point is nicely
illustrated by Fig.~1 of Ref.~\cite{stoks93a}, where the quality of the
$^3$P$_0$ phase shift is shown for an increasing number of short-range
parameters; see also Figs. 2 and 3.) The number of ``counterterms''
needed per partial wave is larger, however, than in Weinberg's power counting
in N$^2$LO. In fact, if we assume that some parameters are needed for
``fine tuning'' to the data ({\em cf.} again Fig.~1 of Ref.~\cite{stoks93a}), or that they
would not be needed if one were fit to the data only up to a lower energy
(say, 250 MeV), then it appears that the number per partial wave is
closer to what Weinberg's power counting would prescribe in N$^3$LO.
It would be very interesting to make this analogy between the Nijmegen
PWA and chiral EFT more precise by fitting the data with a number of
counterterms mandated by the different power-counting schemes.

Another related point, which deserves further attention, is 
that a good description of the $N\!N$ data was obtained by the 
N$^3$LO interactions in Weinberg's counting \cite{entem03a,epelbaum05a}. 
These calculations automatically include the counterterms in all partial 
waves that we consider to be non-perturbative. It is encouraging to 
see that a good description of the data is obtained.
However, lower cutoffs were employed than 
our LO study would suggest. 
This 
surprising fact can be understood if the range of 
cutoffs for which converged results are obtained  increases 
toward lower $\Lambda$ in higher orders of the expansion.
This would be consistent with the observation that the triton binding 
energy is very well described for small cutoffs in Weinberg's NLO. 
This needs to be studied more carefully in the future.

We have taken here the minimalist point of view that only counterterms
that are infrared enhanced with respect to naive dimensional
analysis should be promoted.  
Since we have found no significant cutoff dependence in the $3N$ system in LO,
$3N$ forces are not infrared enhanced at this level, and could
be considered subleading. 
The same is true of the effective range in the $^1$S$_0$ $N\!N$ channel.
For both the $^1$S$_0$ $N\!N$ phase shift and the triton binding energy,
our results are internally consistent. 
However, they are also somewhat unsatisfactory when compared to experiment.
Agreement should be improved in subleading orders.
An alternative, less-conservative approach would be to invoke a promotion
of one or both of these interactions to LO on the basis of fine tuning.
The relatively large value of the effective range
supports this viewpoint.
However, this is largely an issue of convenience that we leave to
later investigation.

The LO results in our power counting are the ones given here.
As we argued, there already exists some evidence that this power 
counting is consistent with previous results. 
A calculation beyond LO is in progress \cite{nogga05a}.

\section{Conclusions}
\label{sec:concl}

In conclusion, we have reanalyzed the predictions of 
chiral perturbation theory in the $N\!N$ system using 
Weinberg's original power counting in LO. We have identified 
that the singularity of the tensor interaction is responsible 
for a significant cutoff dependence in partial waves
where it acts attractively. 
Furthermore, we have 
shown that the addition of 
one 
counterterm in 
each of these partial waves 
removes this cutoff dependence.
It also improves 
the description of the data, which we see as a confirmation 
of our approach. 

For the $3N$ binding energy, we found cutoff-independent results 
in the limit of large cutoffs.
We conclude that the finite 
range of the interaction prevents the Thomas collapse of the 
$3N$ bound state. 
Meaningful predictions for the $3N$ binding energy, 
which are not possible 
using pionless effective field theory, 
are possible here. 
We emphasize that this 
does not invalidate or compromise the pionless EFT approach, but 
reflects the stronger physical constraints built into  ChPT. 

Our approach and its relation to the previous work can be best 
exemplified in the $^3$D$_2$ partial wave. We found that the 
addition of a counterterm ensures 
cutoff independence and increases accuracy. But we also found that in some 
ranges of the cutoff the low-energy description is equally 
good without this additional counterterm. This reconciles the 
traditional approach with our new results. It seems nevertheless 
advisable to promote counterterms in some partial waves. This 
ensures that all partial waves are cutoff independent for the same 
$\Lambda$ and improves the description of the data over a wide range of 
energies. 

We have discussed how these results can be understood from
a power counting that includes angular-momentum suppression.
This improved power counting suggests an ordering of interactions
that is similar to Weinberg's power counting,
except for the infrared enhancement of a few of the counterterms
that contribute to lower partial waves.

Our study 
clearly has to be followed by further investigation in at least two
directions.
First, one would like to understand in more detail 
the interplay of scales in infrared-enhanced counterterms.
A more detailed analysis of the renormalization-group running
and limit-cycle-like behavior of these interactions could shed light on this issue.
Second, one would like to carry out 
a similar investigation for the new NLO and N$^2$LO interactions.
One would like to confirm that a good description of
$N\!N$ data can be obtained already at  N$^2$LO.
Few-nucleon systems should also be reexamined,
as the triton binding energy, for example, does not come out well in LO.  
We consider these to be important remaining 
issues that need to be studied 
for a consistent understanding of the application of 
ChPT to nuclear systems.

\section*{Acknowledgments}
We are grateful to Bruce Barrett, Harald Grie{\ss}hammer,
Hans-Werner Hammer, David Kaplan, and Gautam Rupak 
for enlightening discussions. 
RGET and UvK thank the Department of Physics and the Institute for Nuclear
Theory at the University of Washington for their hospitality during 
part of the period
when this work was carried out.
UvK is also thankful to RIKEN, Brookhaven National Laboratory and 
the U.S. Department of Energy [DE-AC02-98CH10886] for support at
early stages of this work.
The numerical 
calculations have been performed on the IBM SP of the NIC, J\"ulich, Germany.
This work was supported in part by the US DOE (AN, UvK)
and by the 
Alfred P. Sloan Foundation (UvK).

\appendix 

\section{Partial-wave decomposition of OPE}
\label{app:onepi}

The operator form of the OPE potential is 
\begin{equation}
V_{1\pi}(\vec p, {\vec p}\,') 
= - {1 \over (2\pi)^3} \left( { g_A \over 2 f_\pi} \right)^2
\boldtau_1 \cdot \boldtau_2 
{ \vec q \cdot \vec\sigma_1 \ \vec q \cdot \vec\sigma_2 \over \vec q^{\,2} + m_\pi^2 } \ ,
\end{equation}
where $\vec q = \vec p - {\vec p}\,'$ is the momentum transfer.

The isospin operator separates easily. 
It is given by
\begin{equation}
\langle  t  || \boldtau_1 \cdot \boldtau_2 || t' \rangle
=  \left( 2\ t(t+1)-3\right) \delta_{tt'} \ .
\end{equation}
The spin-orbital part can be decomposed into spin and 
orbital tensor operators using obvious basis states,
starting with
\begin{eqnarray}
& & \langle p (ls)j m | \ 
{ \vec q \cdot \vec\sigma_1 \ \vec q \cdot \vec\sigma_2 \over \vec q^{\,2} 
+ m_\pi^2 }
\ | p' (l's')j'm' \rangle \cr
& & = 3 \ 
\langle p (ls)j m | \  
{ \left\{ \sigma_1 \ q \right\}^{00} \ \left\{ \sigma_2 \ q \right\}^{00}  
\over \vec q^{\,2} + m_\pi^2 }
\ | p' (l's')j'm' \rangle \ ,
\nonumber\\
\end{eqnarray}
where we used the representation 
$\vec a \cdot \vec b = -\sqrt{3} \ \left\{ a \ b \right\}^{00}$
of the scalar product of two vectors $\vec a$ and $\vec b$.
With 
$ (\vec q\,)_\lambda  = q \ \sqrt{ 4 \pi \over 3 } \ Y_{1\lambda}(\hat q) $,
we can recouple 
\begin{eqnarray}
& & \langle p (ls)j m | \ 
{ \vec q \cdot \vec\sigma_1 \ \vec q \cdot \vec\sigma_2 \over \vec q^{\,2} 
+ m_\pi^2 }
\ | p' (l's')j'm' \rangle \cr
& & = \sum_{f} { 3 \sqrt{\hat f} \over \sqrt{4 \pi}}  \ 
\left\{ \begin{array}{ccc} 
1 & 1 & f \cr
1 & 1 & f \cr
0 & 0 & 0 \end{array} \right\} \ 
(1\ 1 \ f,00) \ (j' \ 0 \ j,m' \ 0  m ) \cr 
& & \quad  \sqrt{\hat j'} \ 
\left\{ \begin{array}{ccc} 
j & j' & 0 \cr
l & l' & f \cr
s & s' & f \end{array} \right\} \ \sqrt{\hat l \hat s} 
 \langle p \ l \ || \  
{4\pi q^2 \over q^2 + m_\pi^2} \  Y_{f}
(\hat q) \ || \ p' \ l' \ \rangle \cr  
& & \quad  \langle \ s \ || \  
\left\{ \sigma_1 \ \sigma_2 \right\}^{f} 
\ || \ s' \  \rangle \ ,
\end{eqnarray}
thus separating spin and orbital parts. 

The orbital part is 
\begin{eqnarray}
& & \langle p \ l \ m \ | \  
{4\pi q^2 \over q^2 + m_\pi^2} \  Y_{f\mu}
(\hat q) \ | \ p' \ l' \ m' \ \rangle \cr 
& & = \sum_{\lambda_1+\lambda_2=f}
\ \sqrt{ 4 \pi \hat f ! \over \hat \lambda_1 ! \ \hat \lambda_2 !}
\ p^{\lambda_1} \ (-p')^{\lambda_2} \ \sum_k  g_k^f(pp') \cr 
& & \quad \ \sqrt{\hat f} \
\left\{ \begin{array}{ccc} 
k         & k         & 0 \cr
\lambda_1 & \lambda_2 & f \cr
l         & l'        & f \end{array} \right\}  
\ \hat k {\sqrt{\hat \lambda_1 \hat \lambda_2}} \
( k \ \lambda_1 \ l ,00) \ ( k \ \lambda_2 \ l' ,00) \cr 
& & \quad (l' \ f \ l \ , \ m' \ \mu \ m ) \ (-)^{l'} \ \sqrt{ \hat f \over \hat l } \ ,
\end{eqnarray}
where the angular dependence of the propagator was expanded in 
Legendre polynomials using
\begin{equation}
g_k^f(pp') = {\sqrt{\hat k} \over 2 } \ (-)^k
4 \pi \int_{-1}^1 dx \ P_k(x) \ 
{q^2 \over q^2 + m_\pi^2} \  
{ 1 \over q^f } \ .
\end{equation}
This confirms that the 
orbital part is a tensor operator of rank $f$.

With the spin matrix element 
\begin{equation}
\langle \ s \ || \  
\left\{ \sigma_1 \ \sigma_2 \right\}^{f} 
\ || \ s' \  \rangle 
 = 6 \sqrt{\hat {s '} \hat f } \
\left\{ \begin{array}{ccc} 
s         & s'         & f \cr
1/2       & 1/2        & 1 \cr
1/2       & 1/2        & 1 \end{array} \right\} \ ,  
\end{equation}
one obtains the complete matrix elements of the 
OPE:
\begin{eqnarray}
\label{eq:pw1pi}
& & \langle p (ls)j m | \ 
{ \vec q \cdot \vec\sigma_1 \ \vec q \cdot \vec\sigma_2 \over \vec q^{\,2} 
+ m_\pi^2 }
\ | p' (l's')j'm' \rangle \cr
& & \quad = \sum_{f} { 3 \sqrt{\hat f} \over \sqrt{4 \pi}}  \ 
\left\{ \begin{array}{ccc} 
1 & 1 & f \cr
1 & 1 & f \cr
0 & 0 & 0 \end{array} \right\} \ 
(1\ 1 \ f,00) \ (j' \ 0 \ j,m' \ 0  m ) \cr 
& & \qquad  \sqrt{\hat j'} \ 
\left\{ \begin{array}{ccc} 
j & j' & 0 \cr
l & l' & f \cr
s & s' & f \end{array} \right\} \ \sqrt{\hat l \hat s} \cr
& & \qquad \sum_{\lambda_1+\lambda_2=f}
\ \sqrt{ 4 \pi \hat f ! \over \hat \lambda_1 ! \ \hat \lambda_2 !}
\ p^{\lambda_1} \ (-p')^{\lambda_2} \ \sum_k  g_k^f(pp')
\ \sqrt{\hat f} \cr
& & \qquad \left\{ \begin{array}{ccc} 
k         & k         & 0 \cr
\lambda_1 & \lambda_2 & f \cr
l         & l'        & f \end{array} \right\} \ \hat k {\sqrt{\hat \lambda_1 \hat \lambda_2}} \
( k \ \lambda_1 \ l ,00) \ ( k \ \lambda_2 \ l' ,00) \cr 
& & \qquad \ (-)^{l'} \ \sqrt{ \hat f \over \hat l } 
  6 \sqrt{\hat {s '} \hat f } \
\left\{ \begin{array}{ccc} 
s         & s'         & f \cr
1/2       & 1/2        & 1 \cr
1/2       & 1/2        & 1 \end{array} \right\} \ .
\end{eqnarray}

\section{Removal of spurious $NN$ bound states} 
\label{app:tmat}

To remove a spurious bound state in 
the $N\!N$ system we change the interaction $V$ 
to 
\begin{equation}
\bar V = V + | \chi \rangle \ \lambda \ \langle \chi | \ ,
\end{equation}
where $|\chi\rangle$ is the wave function of the spurious bound state,
and $\lambda$ is an energy parameter, which determines a shift of the binding 
energy for the spurious state. The limit $\lambda \to \infty$ removes 
the spurious state, and $\bar V$ is phase-shift equivalent to $V$.

If  
$t$ and $\bar t$  solve 
the LS equations for $V$ and $\bar V$, respectively,
they are related by 
\begin{equation}
\bar t = t + | \eta \rangle \ N \ \langle \eta | \ , 
\end{equation}
with 
\begin{equation}
|\eta \rangle = | \chi \rangle \ + t \ G_0  |  \chi \rangle
\end{equation}
and 
\begin{equation}
N= { \lambda \over 1- \lambda \ \langle \chi | G_0 | \eta\rangle } \ . 
\end{equation}
Here $G_0$ is the free, two-particle Schr\"odinger propagator.
This formulation allows us to perform 
the limit $\lambda \to \infty$ analytically, and we 
end up with the $t$-matrix 
\begin{equation}
\bar t = t - | \eta \rangle \ 
{ 1 \over { \langle \chi | G_0 | \eta\rangle  } } \
\langle \eta | \ ,
\end{equation}
which then enters our calculations for 
the triton binding energy.  
The accuracy of this 
prescription can be checked numerically by 
comparison with the expectation value of the Hamiltonian 
$H$ using the original potential $V$.
 
This procedure can easily be generalized to two and more spurious 
bound states.

\bibliography{lit240505}

\end{document}